\documentclass[aps,prx,longbibliography,superscriptaddress,nofootinbib,twocolumn]{revtex4-2}

\usepackage{dcolumn}

\usepackage{newtxtext}  
\usepackage{bm}         

\usepackage[english]{babel}

\usepackage[pdftex]{graphicx} 
\usepackage{float}
\usepackage{epstopdf}
\usepackage{booktabs}

\usepackage{color}
\definecolor{mygrey}{gray}{0.45}
\definecolor{myblue}{rgb}{0.2,0.2,0.8}
\definecolor{myzard}{cmyk}{0,0,0.05,0}
\definecolor{mywhite}{rgb}{1,1,1}
\definecolor{myred}{rgb}{1,0.,0.3}
\definecolor{kellygreen}{rgb}{0.3, 0.73, 0.09}
\usepackage[colorlinks=true,citecolor=myblue,linkcolor=kellygreen]{hyperref}

\usepackage{latexsym}
\usepackage{amsmath,amssymb,amsfonts,amsbsy,mathtools}
\usepackage{upgreek}    
\usepackage{mathrsfs}   
\usepackage{cancel}     
\usepackage{ulem}       
\usepackage{mathdots}   
\usepackage{setspace}   
\usepackage{dsfont}     

\usepackage{enumerate}  

\usepackage{tikz}
\usetikzlibrary{matrix}

\usepackage{appendix}

\usepackage{natbib}

\DeclarePairedDelimiter\bra{\langle}{\rvert}
\DeclarePairedDelimiter\ket{\lvert}{\rangle}
\DeclarePairedDelimiterX\braket[2]{\langle}{\rangle}{#1 \delimsize\vert #2}

\newcommand\kk{\mathbf{k}}
\newcommand\rr{\mathbf{r}}
\newcommand\RR{\mathbf{R}}

\newcommand\bc[1]{a_{#1}^\dagger}   
\newcommand\bd[1]{a_{#1}}           

\newcommand{\J}[2]{t^{#1}_{#2}}

\newcommand{\dyad}[1]{\overset{\text{\tiny$\bm\leftrightarrow$}}{#1}}

\newcommand\ux{\hat{\mathbf{e}}_x}
\newcommand\uy{\hat{\mathbf{e}}_y}
\newcommand\uz{\hat{\mathbf{e}}_z}

\newcommand\stgmas{m}               
\newcommand\dip{\wp}                

\begin{document}
\title{
Probing and harnessing photonic Fermi arc surface states using light-matter interactions 
} 

\author{I\~naki Garc\'{i}a-Elcano}
\email{innaki.garciae@uam.es}
\affiliation{Departamento de F\'{i}sica Te\'{o}rica de la Materia Condensada  and Condensed Matter Physics Center (IFIMAC), Universidad Aut\'{o}noma de Madrid, E-28049 Madrid, Spain}
\author{Jaime Merino}
\affiliation{Departamento de F\'{i}sica Te\'{o}rica de la Materia Condensada  and Condensed Matter Physics Center (IFIMAC), Universidad Aut\'{o}noma de Madrid, E-28049 Madrid, Spain}
\author{Jorge Bravo-Abad}
\email{jorge.bravo@uam.es}
\affiliation{Departamento de F\'{i}sica Te\'{o}rica de la Materia Condensada  and Condensed Matter Physics Center (IFIMAC), Universidad Aut\'{o}noma de Madrid, E-28049 Madrid, Spain}
\author{Alejandro Gonz\'alez-Tudela}
\email{a.gonzalez.tudela@csic.es}
\affiliation{Instituto de F\'{i}sica Fundamental IFF-CSIC, Calle Serrano 113b, Madrid 28006, Spain}

\begin{abstract}

Fermi arcs, i.e., surface states connecting topologically-distinct Weyl points, represent a paradigmatic manifestation of the topological aspects of Weyl physics. 
Here, we investigate a light-matter interface based on the photonic counterpart of these states and we prove that it can lead to phenomena with no analogue in other setups. First, we show how to image the Fermi arcs by studying the spontaneous decay of one or many emitters coupled to the system's border. Second, we demonstrate that the Fermi arc surface states can act as a robust quantum link. To do that we exploit the negative refraction experienced by these modes at the hinges of the system. Thanks to this mechanism a circulatory photonic current is created which, depending on the occurrence of revivals, yields two distinct regimes. In the absence of revivals, the surface states behave as a dissipative chiral quantum channel enabling, e.g., perfect quantum state transfer. In the presence of revivals, an effective off-resonant cavity is induced, which leads to coherent emitter couplings that can entangle them maximally.
In addition to their fundamental interest, our findings evidence the potential offered by the \textit{photonic Fermi arc light-matter interfaces} for the design of more robust quantum technologies.

\end{abstract}

\maketitle

\section{Introduction} 
The introduction of topology to explain the observation of quantized electron transport~\cite{Klitzing1980} has led to a revolution in Physics, permeating in fields beyond condensed-matter, such as photonics~\cite{Ozawa2019} or acoustics~\cite{Huber2016,Ma2019}. On the fundamental side, it has brought the discovery that certain phases of matter can only be characterized by global order parameters \cite{Ryu2010,Qi2011,Chiu2016,Wen2017}, escaping thus to the Ginzburg-Landau paradigm. From a more applied standpoint, such topological phases are accompanied by the appearance of topological boundary states. Owing to their topological origin, these boundary states are immune to disorder and thus can be used to engineer robust devices. Initially, the field focused on topological phases and their boundary states in one and two-dimensions, such as 2D Chern ($\mathbb{Z}_2$) insulators and their chiral~\cite{Haldane1988} (helical) edge modes~\cite{Kane2005a}. However, the observation in 2008 of the first 3D topological insulator~\cite{Hsieh2008}, and in 2015 of Weyl semimetals in electronic~\cite{Xu2015,Lv2015} and photonic~\cite{Lu2015} setups has driven the attention to the three-dimensional case~\cite{Hasan2010,Armitage2018a}.

Weyl systems, in particular, stand as one of the most paradigmatic examples of a three-dimensional topological phase. They are characterized by the presence of several single-point linear degeneracies in their bulk spectrum, known as Weyl points, which have associated a quantized Berry curvature. Such quantization triggers the appearance of topological surface states with an energy dispersion connecting two topologically inequivalent Weyl points: the Fermi arcs~\cite{Wan2011a,Haldane2014}. These unconventional surface modes are responsible of exotic phenomena in electronic systems such as bulk-mediated quantum oscillations~\cite{Potter2014,Moll2016,Wang2017a,Zhang2019}, as well as unusual classical wave propagation in bosonic settings such as photonics or acoustics~\cite{He2018a,Yang2019,Deng2020,Cheng2020,Guo2021,Li2022,Wang2022}. However, an important practical difference between the two scenarios is that whereas the electronic ground state fills up until the Fermi level, and thus naturally probe Fermi arc energies, bosonic excitations accumulate in the lowest energy state due to their statistics. Thus, the phenomena and detection schemes introduced in the electronic context are not, in general, directly extrapolable to the bosonic ones~\cite{Atala2013,Mittal2014,Aidelsburger2015,Hu2015,Zeuner2015,Duca2015,Mittal2016,Li2016,Wimmer2017,Cardano2017,Sugawa2018,Elben2020,Jiao2021,Leykam2021}.  

In Photonics, this limitation, which is a commonplace for all topological models, is motivating interfacing such structures with emitters~\cite{Barik2018a,Mehrabad2020,Kim2021,Owens2021}. The reason is that emitters can probe the photonic system at topologically non-trivial frequency regions, making them active. As an added value, emitters are strongly interacting systems that can induce photonic interactions through light-matter couplings. These recent experimental developments are driving many theoretical studies, which, so far, have focused on understanding the photon-mediated interactions when emitters couple to topological bulk modes~\cite{Bello2019,Bello2022,Leonforte2021a,Vega2021,Kim2021,DeBernardis2021,Garcia-Elcano2020,Garcia-Elcano2021,BlancodePaz2022a}. Recent studies with one~\cite{Almeida2016} and two-dimensional topological photonic systems~\cite{Vega2022} have discovered how the interaction of emitters with the topological boundary modes can also lead to novel regimes in cavity and waveguide QED. Remarkably, the interaction of topological surface states with emitters has yet not been studied, and thus their potential applications remain still an open question.

In this work, we undertake this endeavour by characterizing a \textit{Fermi arc light-matter interface}, consisting of a set of emitters coupled to the edges of a Weyl system, and find several unexpected phenomena. First, we demonstrate that one can directly image the Fermi arcs by monitoring the free-space spontaneous emission of the emitters. 
Second, we prove that the surface modes can act as a robust quantum link connecting the emitters in two different ways. 
In the infinite size limit, we show how to engineer the negative refraction occurring at the hinges of the system to obtain a perfect, dissipative, chiral channel~\cite{Lodahl2017}, and characterize its performance by studying the spontaneous formation of entanglement. Interestingly, we demonstrate that the studied channel can reach the maximum entangling capacity of a perfect chiral waveguide~\cite{Gonzalez-Ballestero2015}, which opens up its use for quantum state transfer~\cite{Cirac1997} or to obtain driven-dissipative maximally entangled states~\cite{Ramos2014a,Pichler2015a,Ramos2016a}. 
For small systems, meaning that revivals of the emitters occur, the multiple negative refractions taking place at corners of the structure facilitate the formation a closed photonic path which induces an effective cavity mode. Such cavity leads to perfect coherent exchanges between the emitters, which can be used to maximally entangle them in a time-dependent fashion.

The manuscript is structured as follows: in Section~\ref{sec:fermilm}, we describe the Fermi arc light-matter interface that we consider along the manuscript, with special emphasis on the properties of the topological surface states; in Section~\ref{sec:probing}, we explain how to image the Fermi arcs by monitoring the emitters's spontaneous emission; in Section~\ref{sec:harnessing}, we explain how to harness the negative refraction phenomena occurring at the system's hinges to engineer perfect coherent and dissipative quantum links with the topological surface modes of the structure; and finally, in Section~\ref{sec:conclu}, we end our manuscript by summarizing our main findings.

 \begin{figure*}[t]
   \centering
    \includegraphics[width=17.8cm]{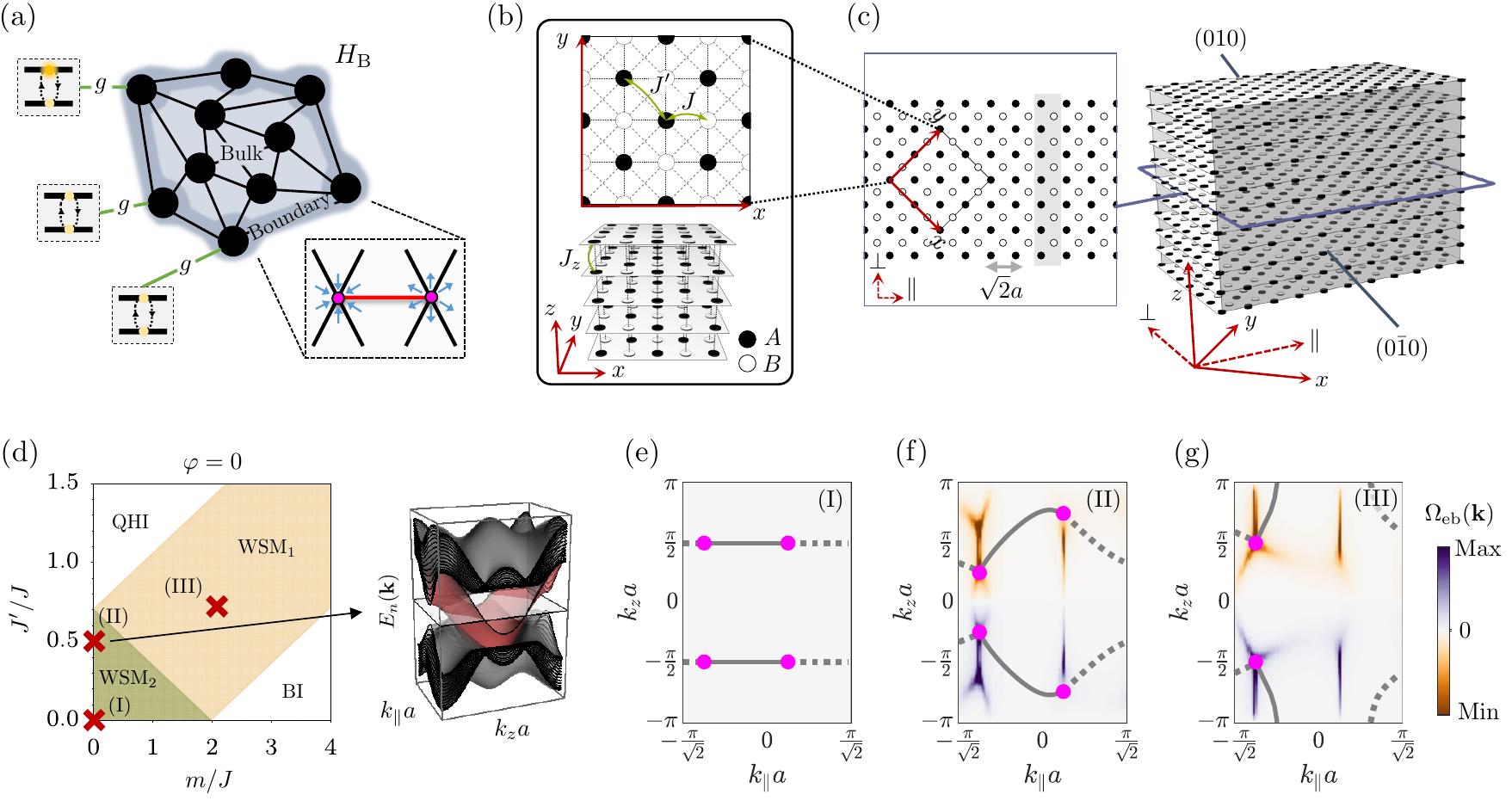}
       \caption{
       {\bf Photonic Weyl environment realized in a three-dimensional lattice of localized bosonic modes.} 
       (a)~Schematic view  of the investigated system. 
       (b)~Hopping amplitudes associated to the intra-layer (top panel) and inter-layer (bottom panel) interactions in the discrete lattice model.  Black and white shallow cylinders represent the localized bosonic modes belonging to the $A$ and $B$ sublattices, respectively.
       (c)~Slab model for the considered cut. Left panel shows the $z=0$ plane of the photonic lattice, where the shadowed grey area represents the unit cell associated to the cut system. Right panel renders a three-dimensional view of the slab. As seen, we choose both the $(0\bar{1}0)$ and $(010)$ facets to be composed by sites belonging to the $A$ sublattice.
       (d)~Phase diagram of the Weyl photonic environment setting $\varphi=0$ [see definition in Eq.~\eqref{eq:d(k)}], as a function of the on-site energy off-set between sublattices $m$ and the next-nearest neighbors hopping elements $J^{\prime}$. Inset shows the band structure corresponding to the (II) configuration, where the red surface represent the edge band. 
       (e-g)~Fermi arcs associated to three different configurations marked by red crosses in the phase diagram displayed in (d). Solid and dotted lines differentiate between states associated to the $(0\bar{1}0)$ and the $(010)$ facets, respectively. Magenta points denote the projection of the Weyl singularities over the surface Brillouin zone. The color map indicates the Berry curvature calculated for the edge band $\Omega_{\mathrm{eb}}(\kk)$, in each of the considered cases. For the calculation, a slab of width $w/a=16\sqrt{2}$ (i.e., 33 non-equivalent sites per unit cell) is employed. Therefore, the edge band is identified as the $n=17$ band. Hot lines along which the surface Berry curvature displays a non-trivial behaviour are present in panels (f) and (g).
       }
 \label{fig:Fig1}
 \end{figure*}

\section{Fermi arc light-matter interfaces~\label{sec:fermilm}}

In this Section, we provide all the details about the Fermi arc light-matter interface that we will consider along this manuscript. First, we briefly describe the light-matter Hamiltonians that we use in Section~\ref{subsec:lm}. Then, in section~\ref{subsec:Weylbath}, we make an extensive description of the considered photonic Weyl environment, explaining its bulk and boundary properties.

\subsection{Light-matter coupling scheme~\label{subsec:lm}}

A schematic overview of the system under study is depicted in Fig.~\ref{fig:Fig1}(a). 
We consider one or more emitters coupled to the boundary of a photonic Weyl environment, which we model as a discrete photonic lattice. The interplay among the modes spanning such topological reservoir is captured by a tight-binding Hamiltonian of the form (taking $\hbar=1$ hereafter):
\begin{equation}
    H_\mathrm{B} = \sum_{\rr\rr^{\prime}}J_{\rr\rr^{\prime}}\bc{\rr}\bd{\rr^{\prime}},
\end{equation}
where $a_{\rr}(a_{\rr}^{\dagger})$ annihilates (creates) a bosonic excitation at position $\rr$ and $J_{\rr\rr^{\prime}}$ is, in the most general case, a complex hopping matrix element. 
The band structure resulting from the diagonalization of $H_\mathrm{B}$ harbours an even number of Weyl points. These are point-like degeneracies in reciprocal space around which, assuming that we limit ourselves to the study of the so-called type I Weyl semimetals with no tilting velocity terms~\cite{Soluyanov2015,Armitage2018a}, the dispersion can be expanded as follows:
\begin{equation}
\label{eq:Weyl_dispersion}
     E_{\pm}(\kk\sim\kk_W^{i}) \approx \omega_W\pm
     \sqrt{\sum_{\alpha\alpha^\prime=x,y,z}
     \bar{M}^{i}_{\alpha\alpha^\prime}\,q_\alpha^{i} q_{\alpha^\prime}^{i}},
\end{equation}
where we define $\omega_W$ as the Weyl frequency, $\mathbf{q}^{i}=\kk-\kk_W^{i}$ denotes the distance between an arbitrary point in the Brillouin zone $\kk$ and the position of the $i$-th Weyl node $\kk_W^{i}$, and $\bar{M}^{i}$ is a positive definite matrix. 

For the emitters we use the simplest description, that is, considering them as two-level systems with resonant frequency $\omega_j$. The emitters' operators will be denoted by $\sigma_{\nu\nu^\prime}^j=\ket{\nu}_j\bra{\nu^\prime}_j$ $\left(\nu,\nu^\prime=g,e\right)$, where $\ket{e}_j$ and $\ket{g}_j$ stand for the excited and ground state of the $j$-th emitter, respectively. 
Then, provided that the emitters are locally coupled to specific sites of the photonic bath $\rr_j$, the Hamiltonian of the full system reads as [see Fig.~\ref{fig:Fig1}(a)]:
\begin{equation}
    H = H_\mathrm{B} + \sum_{j=1}^{N}\left(\omega_{j}\sigma_{ee}^{j} +
    \sum_\rr g_{\rr\rr_j}\,\bc{\rr}\sigma_{ge}^{j}+\mathrm{H.c.}\right),
\end{equation}
where $N$ is the total number of emitters and $g_{\rr\rr_j}=g\,\delta_{\rr\rr_j}$, with $g$ representing the light-matter coupling strength. This type of light-matter coupling Hamiltonians can describe both the situation where natural or artificial atoms couple to real photonic crystal environments~\cite{Lu2015} and other where superconducting qubits couple to microwave resonator arrays~\cite{Liu2017,Mirhosseini2018,Sundaresan2019,Kim2021}. Besides, it is noteworthy that similar light-matter interaction Hamiltonians can be emulated with purely atomic setups by replacing the role of photons by matter-waves~\cite{DeVega2008,Navarrete-Benlloch2011a}. The latter is a particularly promising platform to test our predictions given that both the first implementation of such simulated light-matter interfaces~\cite{Krinner2018} and Weyl points~\cite{Wang2021} has been recently achieved.

\subsection{Tailoring the Weyl environment~\label{subsec:Weylbath}}

The discrete lattice model that embodies the photonic Weyl environment is given, for definiteness, by a generalization of the proposal described in Ref.~\cite{Dubcek2015a}, which we design to break inversion and time-reversal symmetries. However, we expect that the conclusions we derive from this model can be extended to any tight-binding scheme featuring similar dispersive properties, particularly to those hosting a prototypical type I semimetallic phase~\cite{Delplace2012a,Hou2016a,Goikoetxea2020a}.

A real space representation of the considered model is outlined in Fig.~\ref{fig:Fig1}(b). It is useful to present it as a set of bidimensional layers, consisting of square lattices, stacked along the $z$-axis. The sites within each layer are connected through nearest- and next-nearest-neighbors interactions (with amplitude $J$ and $J'$, respectively), whereas inter-layer couplings occurs solely between vertically aligned sites (with amplitude $J_z=J$). 
To obtain the Weyl phase we impose a non-trivial phase pattern which involves the definition of a two-sites unit cell. The later introduces a sublattice degree of freedom that acts as a ``pseudospin''. In Figs.~\ref{fig:Fig1}(b) and (c), the sites of what we define as the $A$ and $B$ sublattices are represented by black and white shallow cylinders, respectively. Finally, we include a staggered mass term $m$ which creates an onsite energy off-set between the modes belonging to the two different sublattices.
Assuming periodic boundary conditions, the Hamiltonian matrix can be written in reciprocal space as follows (see details in Appendix~\ref{apx:lattice}): 
\begin{equation}
\label{eq:Bloch_Hamiltonian}
    \bar{H}_\mathrm{B}(\kk)=\omega_W\mathds{1}+\mathbf{d(\kk)}\bm{\sigma},
\end{equation}
where $\bm{\sigma}=(\sigma_x,\sigma_y,\sigma_z)$, with $\sigma_{x,y,z}$ representing the Pauli matrices, and $\mathbf{d(\kk)}$ is a $\kk$-dependent vector whose components are given by:
\begin{equation}
\label{eq:d(k)}
    \begin{cases}
    d_x(\kk) = -J\left[\cos(k_xa+\varphi)+\sin(k_xa)+2\cos(k_ya)\right],
    \\[0.5ex]
    d_y(\kk) = +J\left[\sin(k_xa+\varphi)+\cos(k_xa)\right],
    \\[0.5ex]
    d_z(\kk) = -m - 2J\cos(k_za) + 4J^{\prime}\sin(k_xa)\sin(k_ya).
\end{cases}
\end{equation}
Here, $a$ denotes the distance between nearest neighbours and we define $\varphi$ as the complex phase picked up by the excitation when it jumps from a $B$ site to an $A$ site in the negative $x$ direction. In the following, to center the discussion, we fix $\varphi=0$ unless explicitly specified. 

\textbf{\textit{Phase diagram}}.
The positions of the Weyl points in the Brillouin zone are obtained by solving $|\mathbf{d(\kk)}|=0$. In order to delineate the different regions that comprehend the phase diagram shown in Fig.~\ref{fig:Fig1}(d), we study the moving and merging of these Weyl nodes as we sweep the parameter's space spanned by $m$ and $J^\prime$. The green shadow area occupying the lower left corner of the phase diagram corresponds to a Weyl semimetallic phase featuring two pairs of Weyl points ($\mathrm{WSM_{2}}$). This phase can evolve into a different type of Weyl semimetal which is characterized by the presence of a single pair of band touching points ($\mathrm{WSM_{1}}$).
Interestingly, the $\mathrm{WSM_{1}}$ phase emerges in between a normal insulator ($\mathrm{BI}$) and a quantum (anomalous) Hall insulator ($\mathrm{QHI}$), as expected~\cite{Murakami2007}. The topological characterization of these gapped phases can be done using a dimensional reduction strategy~\cite{Jiang2012,Delplace2012a}. The latter consists in calculating the Chern number of an effective 2D model stemming from treating $k_z$ as a free parameter in the Bloch Hamiltonian defined by Eq.~\eqref{eq:Bloch_Hamiltonian}. We have found that the topological invariant of the $\mathrm{QHI}$ phase is $C=\pm1$ for all values of $k_z$, whereas the Chern number in the $\mathrm{BI}$ phase is always zero (see details in Appendix \ref{apx:topo_phases}).

\textbf{\textit{Emergence of the Fermi arcs}}.
As a natural consequence of the bulk-edge correspondence, when the photonic environment is prepared in the Weyl semimetallic phase, we expect topologically protected edge states to show up. 
In order to characterize the surface modes of our model, a slab-like geometry is investigated. To implement that, we impose open boundary conditions along some specific spatial direction while preserving the discrete translation symmetry within the remaining ones. 
For the sake of conciseness, we consider a slab featuring an infinite size along the $z$ and $x+y$ directions, but a finite width along the $x-y$ direction, as shown in Fig.~\ref{fig:Fig1}(c). The boundaries of such slab correspond to two bi-dimensional planes, which we choose to be composed by sites belonging to the $A$ sublattice and that we refer to as the $(010)$ and the $(0\bar{1}0)$ facets.
Note that, for convenience, we have introduced a coordinate system rotated $\pi/4$ around the $z$-axis. In this new basis, the lattice vectors are $\mathbf{a}_1=\sqrt{2}a\,\mathbf{\hat{e}}_\parallel$, $\mathbf{a}_2=\sqrt{2}a\,\mathbf{\hat{e}}_\perp$ and $\mathbf{a}_3=a\,\mathbf{\hat{e}}_z$, where $\mathbf{\hat{e}}_{\genfrac{}{}{0pt}{4}{\parallel}{\perp}}=\frac{1}{\sqrt{2}}(\mathbf{\hat{e}}_x\pm\mathbf{\hat{e}}_y)$ and $\mathbf{\hat{e}}_\alpha$ denotes a unit vector in the $\alpha$ direction.

The periodicity of the lattice along the $z$ and $\parallel$ directions can be exploited to exactly diagonalize the bath Hamiltonian with the selected boundary conditions. To do so, we consider an extended unit cell [see the gray shadowed area in the left panel of Fig.~\ref{fig:Fig1}(c)] and we define the surface lattice vectors as $\mathbf{a}_1^{s}\equiv\mathbf{a}_1$ and $\mathbf{a}_2^{s}\equiv\mathbf{a}_3$. Equipped with these two elements, one can invoke the Bloch theorem to bring $H_\mathrm{B}$ to its diagonal form.
The $\kk$-space's primitive cell of this quasi-two-dimensional  model is precisely the surface Brillouin zone. There, the Fermi arcs appear as the $E_n(\kk)=\omega_W$ equifrequency contours of the calculated band structure [see inset of Fig.~\ref{fig:Fig1}(d)], where $E_n(\kk)$ denotes the $n$-th band. 

The Fermi arcs associated to the $(010)$ (gray dotted lines) and the $(0\bar{1}0)$ (gray solid lines) facets are shown in Figs.~\ref{fig:Fig1}(e-g) for three different points of the phase diagram. As seen, they stand as open curves that connect the projections over the surface Brillouin zone of Weyl points with different chirality. The displayed color maps depict the surface Berry curvature associated to the edge band $\Omega_\mathrm{eb}(\kk)$, which corresponds to the red colored band in the inset of Fig.~\ref{fig:Fig1}(d). These maps show that, for the (II) and (III) configurations, ``hot lines'' along which the surface Berry curvature exhibits a divergent behaviour emerge~\cite{Wawrzik2021}. These are associated to areas in which the localization of the wavefunction changes drastically (see details in Appendix \ref{apx:slabBC}).

 \begin{figure*}[t]
   \centering
    \includegraphics[width=17.8cm]{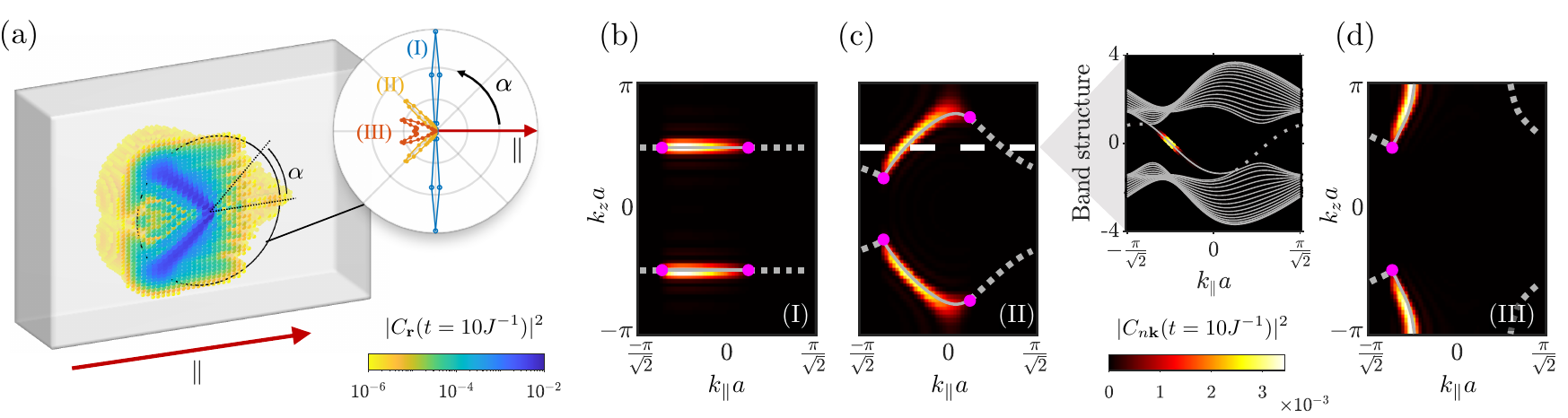}
     \vspace{0.0cm}
       \caption{{\bf Probing the surface modes of the Weyl system via local light-matter coupling with a single emitter.} (a) Photonic component of the overall-system wave function at $tJ=10$ for the configuration marked by (II) in the phase diagram of Fig.~\ref{fig:Fig1}(d). The system is prepared with the emitter in its excited state and couple to the central site of the $(0\bar{1}0)$ surface. Inset shows the probability of finding the photonic excitation at $\rr=\mathcal{R}(\cos\alpha,\sin\alpha)$, with $\mathcal{R}/a\approx20$, after a measuring time of $tJ=60$, for each of the three cases considered in the phase diagram of Fig.~\ref{fig:Fig1}(d). (b-d) Distribution of the Bloch modes' population at $tJ=10$ for the three considered cases. In all the calculations, the light-mater coupling strength is $g/J=0.5$. A finite slab of width $w/a=16\sqrt{2}$ is employed. The dimensions of the $(0\bar{1}0)$ facet in the $\parallel$ and $z$ directons are $L_\parallel/a=63\sqrt{2}$ and $L_z/a=63$, respectively. To compute the inset, of panel (a) a slightly larger slab is used to avoid reflection effects with the system's edges.
       }
 \label{fig:Fig2}
 \vspace{-0.3cm}
 \end{figure*}
 
\section{Probing Fermi arc surface states via spontaneous emission~\label{sec:probing}}

As mentioned in the introduction, the absence of a Fermi energy in the photonic setting makes the detection of the Fermi arcs not directly extrapolable from the electronic context. The first measurements used angle-resolved transmission to detect the Fermi arcs in photonic systems~\cite{Lu2015,Chen2016}, whereas more refined experiments using classical local probes and near-field scanning measurements~\cite{Noh2017,Yang2017,Yang2018,Li2018} have been able to provide better visualization of these modes. In this Section, we show that Fermi arc light-matter interfaces represent an outstanding alternative way of probing and imaging the Fermi arcs by monitoring emitter's spontaneous emission. For that, we first show in section~\ref{subsec:single} that a single emitter locally coupled to the edge of a Weyl system naturally excites surface modes into the bath. Then, in section~\ref{subsec:many} we show that if one couples not one but many emitters, and monitor their spontaneous emission far from the surface, one can have a visualization of the Fermi arcs of the bath they are coupled to.

\subsection{Launching Fermi arc surface modes with a single emitter~\label{subsec:single}}

Let us start by considering a single emitter, prepared in its excited state, and locally coupled to one of the sites in the lattice's boundary. The photonic excitation is injected in the system via spontaneous emission and its propagation through the reservoir can be studied using an exact treatment. The later implies solving the Schr\"{o}dinger equation for large finite baths~\cite{Garcia-Elcano2021}. 
For that, we introduce an overall-wavefunction ansatz of the form (setting the emitter's number $N=1$):
\begin{equation} \label{eq:overall_wavefun}
    \ket{\Psi(t)}=\left[
    \sum_{j=1}^N C_{j}(t)\sigma_{eg}^j+
    \sum_{\rr}C_{\rr}(t)\bc{\rr}
    \right]\ket{\Psi_\mathrm{vac}},
\end{equation}
where $|C_j(t)|^2$ and $|C_\rr(t)|^2$ yield the population of the $j$-th emitter and of the bosonic mode localized at position $\rr$, respectively, and $\ket{\Psi_\mathrm{vac}}\equiv\ket{g_1\cdots g_N;\mathrm{vac}}$, with $\ket{\mathrm{vac}}$ the electromagnetic vacuum.
Since our calculations are performed in a finite system, a specification of the photonic lattice's shape is needed. We consider the one shown in Fig.~\ref{fig:Fig2}(a), which resembles (modulo its finite size) the slab described in the previous section. 
Then, the emitter, which we locate in the center of the $(0\bar{1}0)$ surface, is tuned to the Weyl frequency.

Under free evolution, the emitter relaxes to its ground state and the photonic excitation is transferred to the Weyl environment. {Due to the local character of the light-matter coupling, the emitter excites preferentially all the $\kk$-modes resonant to its energy. Since we have chosen the emitter's frequency to lie at the Weyl frequency}, it will couple mostly to the surface modes, and thus the excitation spreads out,  mostly, among the sites conforming the boundary wherein the emitter is placed. In Fig.~\ref{fig:Fig2}(a), we show the photonic population at each lattice site for the temporal frame $tJ=10$, with $H_\mathrm{B}$ prepared in the configuration marked by $(\mathrm{II})$ on the phase diagram of Fig.~\ref{fig:Fig1}(c) and assuming that $g/J=0.5$. There, the presence of two channels of highly collimated emission oriented in the upward and downward directions reflects the mirror symmetry displayed by the Weyl bath along the $z=0$ plane. Besides, we observe that the emission presents a strong chiral behaviour featuring a V-shaped emission profile with no analogue in locally coupled light-matter interfaces~\cite{Gonzalez-Tudela2019}. The yellow line in the inset of Fig.~\ref{fig:Fig2}(a) shows the probability of finding the excitation at a given angle $\alpha$ within a circle of radius $\mathcal{R}/a\approx20$, centered at the emitter’s position, after a measuring time of $tJ=60$. This calculation is repeated for configurations (I) and (III) yielding the blue and orange plots, respectively, showing how one can control the emission patterns through the bath parameters. 

If we restrict the study of the system's dynamics to time values such that $tv<L$, where $v\sim Ja$ is the average velocity at which the excitation propagates through the bath and $L$ accounts for the linear extent of the surface wherein the emitter is placed, reflection effects at the facet's borders can be neglected. In that case, the results stemming from solving the Schr\"{o}dinger equation in the finite-sized lattice do not differ from those that one would have obtained if an infinite slab had been considered. Thus, it is legitimate to introduce an alternative representation in which the evolution of the photonic component is described by the time evolving population of the ensemble of Bloch modes that diagonalize $H_\mathrm{B}$ when the slab-like boundary conditions are imposed (see Appendix \ref{apx:mapping}). The time-dependent population of each Bloch mode can be calculated as follows:
\begin{equation}
    |C_{n\kk}(t)|^2=|\braket{\psi_{n\kk}}{\Psi(t)}|^2
\end{equation}
where $\ket{\psi_{n\kk}}$ is the Bloch state associated to the $n$-th band and the quasi-momentum $\kk=(k_\parallel,k_z)$. Pretty much in the same way as in the electronic context, one can build some intuition upon the excitation's dynamics by adopting a semiclassical description~\cite{Xiao2015}. Within this picture one can relate the speed and direction of the propagating excitation to the group velocities of the Bloch modes that couple to the emitter. 

The population of Bloch modes arising from the photonic profile presented in Fig.~\ref{fig:Fig2}(a) is shown in Fig.~\ref{fig:Fig2}(c). Together with that, we plot the projections of the Weyl points over the surface Brillouin zone (magenta dots).
We observe that the excited modes outline the shape of the Fermi arcs corresponding to the surface in which the emitter is placed (see the straight gray lines highlighted in the main panel). The inset displays a cut of the band structure for $k_za=\pi/2$, where the colored points reveal the energy distribution of the excited modes. Similar treatment is performed for the configurations $(\mathrm{I})$ and $(\mathrm{III})$, obtaining equivalent mappings as demonstrated in Figs.~\ref{fig:Fig2}(b) and (d), respectively.
{
In order to understand why this imaging occurs, we first note that, as long as we remain in the regime where $g/J<1$, the modes playing the most relevant role in the dynamical process are those in resonance with the emitter's frequency. In our case, as we are considering a type I Weyl semimetal and tuning the emitter to the Weyl frequency, we can ensure that the dynamics will be dominated by the Bloch states comprising the Fermi-arcs, namely, because they are the ones whose associated energy coincides with $\omega_W$. Besides, the fact that the emitter acts as a local probe in real space translates into a strong non-local character of the coupling in $\kk$-space, which means that all the modes conforming the Fermi arcs are, in principle, homogeneously excited. Nonetheless we must take into account that, since leakage of the photonic population to the bulk is strongly suppressed whenever the emitter is in resonance with the Weyl frequency, only the Fermi arcs corresponding to the specific boundary where the emitter is embedded into can be actually addressed.
}

\begin{figure}[t]
   \centering
    \includegraphics[width=\columnwidth]{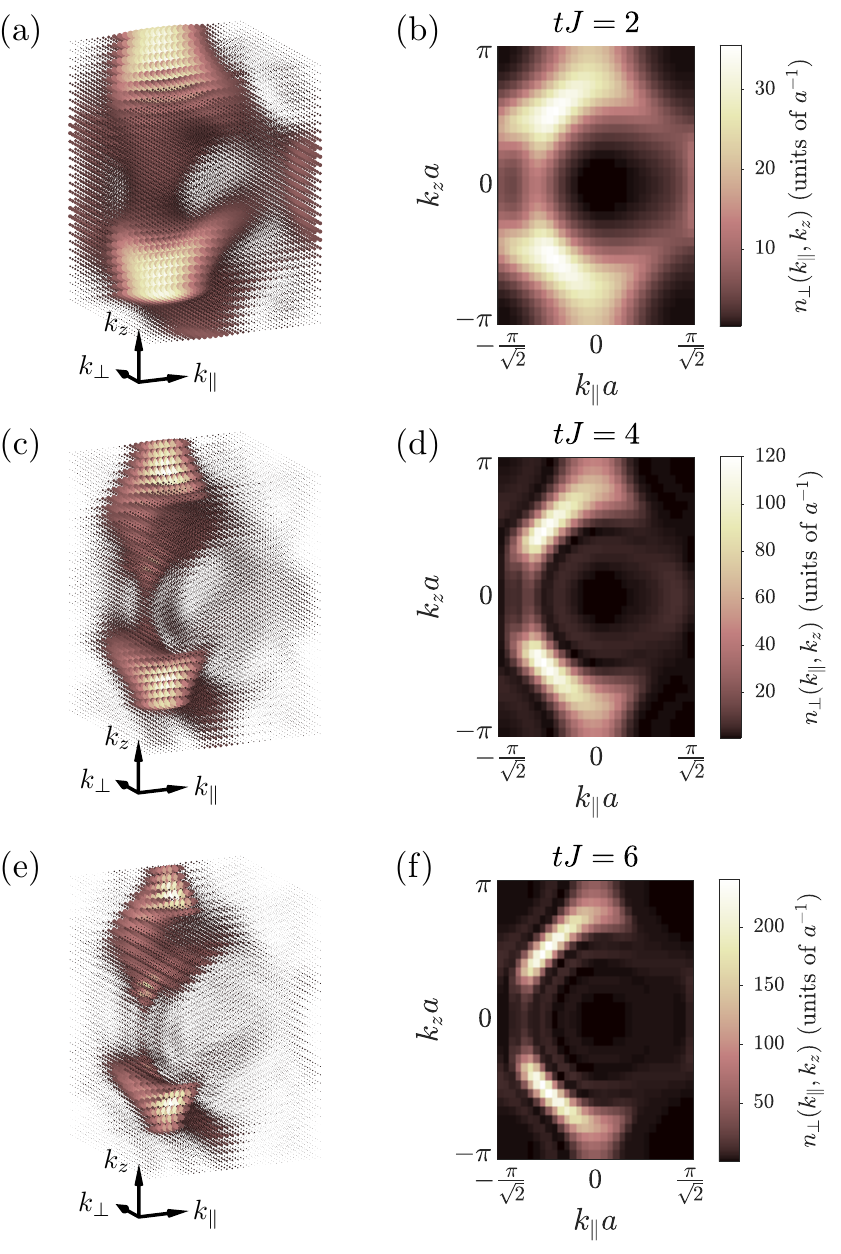}
       \caption{ 
       {\bf Fermi arcs visualization via time-of-flight imaging of the Weyl photonic environment coupled to an initially excited emitter}.
       We consider a finite lattice conformed by 31 sites in the $\parallel$ and $\perp$ directions and 61 sites in the $z$ direction [see Fig.\ref{fig:Fig1}(c)], prepared in the configuration marked by (II) in the phase diagram of Fig.\ref{fig:Fig1}(d), and harbouring an initially excited emitter coupled to the center of the $(0\bar{1}0)$ facet, with $g/J=0.5$.
       (a,c,e) Three-dimensional momentum distribution $n(\kk)$ of the photonic modes excited by the emitter for released times of $tJ=2$, $4$, and $6$, respectively. 
       (b,d,f) Projection of the momentum distribution obtained in (a,c,e) along the $k_\perp$ direction. 
       }
 \label{fig:Fig3}
 \end{figure}

\textit{Transforming from real to reciprocal space representation using time-of-flight pictures.}
The dynamical behaviour illustrated in Fig.~\ref{fig:Fig2}(a) shows that the emitter is acting as a local probe launching surface modes over the photonic structure~\cite{Chen2016,Noh2017,Li2018,He2018a}. A natural question to ask is whether there is a way of recovering the ``Fermi arc picture'' once the excitations are transfer into the bath degrees of freedom. Here, we want to illustrate that, for the case of the emulated light-matter interfaces using ultra-cold atoms~\cite{DeVega2008,Navarrete-Benlloch2011a,Krinner2018,Stewart2020}, in which the photons are nothing more than matter waves propagating through an optical lattice, there is a straightforward way of doing it. The idea consists in, once the excitation has been launched, remove the optical traps so that the matter-waves are released. This is the method known as time-of-flight imaging~\cite{Greiner2002}, and it is used routinely in cold atom experiments~\cite{Bloch2005}. After switching off the trap, the density distribution of the propagating atomic cloud can be shown to be related to the following momentum distribution (see Appendix \ref{apx:time_of_flight}):
\begin{equation} 
    n(\kk)=\sum_{j,j^\prime}e^{i\kk(\rr_j-\rr_{j^\prime})}\bra{\Psi(t)}a^\dagger_{\rr_j} a_{\rr_{j^\prime}}\ket{\Psi(t)}\,,
\end{equation}
which can be measured through resonant absorption imaging after releasing the matter-waves. In the left column of Fig.~\ref{fig:Fig3} we plot the 3D momentum distribution $n(\kk)$ associated to the photonic population featured by the Weyl environment, prepared in the configuration marked by (II) in Fig.~\ref{fig:Fig1}(c), for several de-excitation times in the different rows, whereas in the left column we plot the corresponding the column-integrated momentum distribution along the $\perp$ direction: $n_\perp(k_\parallel,k_z)=\int dk_\perp\,n(\kk)$.
In these panels, we observe that, as the emitter excitation decays completely to the environment, the Fermi arc shape found using the formal mapping defined in the previous section emerges [compare Fig.~\ref{fig:Fig3}(f) with Fig.~\ref{fig:Fig2}(c)].

\subsection{Fermi arc imaging through many emitters' spontaneous emission~\label{subsec:many}}

\begin{figure}[t]
   \centering
    \includegraphics[width=\columnwidth]{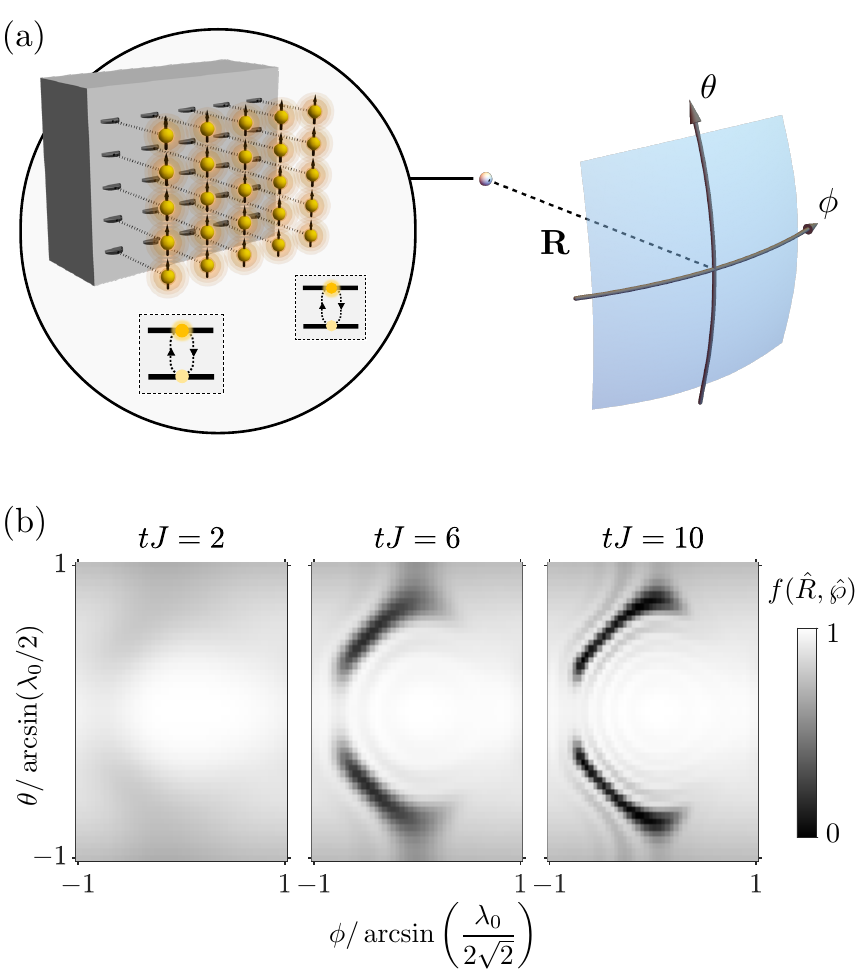}
       \caption{{\bf Radiation emanating from an emitter array coupled to the boundary of the Weyl enviornment}. 
       (a) Sketch of the considered situation: the radiation stemming from an array of emitters, locally coupled to the sites conforming the $(0\bar{1}0)$ facet of the Weyl lattice [see  Fig.\ref{fig:Fig1}(c)], is collected in the far-field as a function of the angular variables $\theta$ and $\phi$. The light-matter coupling strength $g$, polarization direction $\hat{\dip}$, and transition's wavelength $\lambda_0$ are assumed to be the same for all the considered emitters. At the initial time only the emitter placed in the center of the face is in its excited states, with no photonic excitations in the bath.
       (b) Angular dependence of the radiation pattern obtained for three three different time frames provided that the Weyl bath is prepared in the configuration marked by (II) in the phase diagram of Fig.\ref{fig:Fig1}(d) and that $g/J=0.2$, $\hat{\dip}=\hat{z}$, and $\lambda_0/a=1$.
        }
 \label{fig:Fig4}
 \end{figure}

A final question regarding the imaging of Fermi arcs is whether there is a method that does not rely on the matter-wave nature of the photonic excitation. In this section, we provide a positive answer by showing that, by monitoring the free-space spontaneous emission of an emitter array attached to the border of the Weyl environment, one can image the Fermi arcs in reciprocal space. To illustrate that, we consider an array of emitters coupled to the $(0\bar{1}0)$ surface (one emitter per lattice site in the facet), as shown in Fig.~\ref{fig:Fig4}(a). Then, we assume that the central emitter is excited while the rest are in their ground state. One can show that as the central emitter decays into the bath as surface modes, it will also excite the rest of the atoms, that will eventually decay into the bath as well. Interestingly, if one consider that the emitters not only decay into the bath, but also radiate into free-space modes one can monitor the formation of the Fermi arcs in real space. In particular, if we assume that the emitters radiate as electric dipoles, the intensity of the light being emitted at position $\RR=|\RR|\hat{R}$ is given by $\bra{\Psi(t)}\mathbf{E}^{-}(\mathbf{R})\mathbf{E}^{+}(\mathbf{R})\ket{\Psi(t)}$. Here, $\mathbf{E}^{+}(\RR)$ stands for the positive frequency component of the electric field that, in the Markovian approximation, reads~\cite{Asenjo-Garcia2017}:
\begin{equation} \label{eq:E_plus(R)}
    \mathbf{E}^{+}(\RR) = 
    \mu_0\omega_0^2\sum_{j=1}^{N}
    \dyad{G}_0(\RR,\rr_j,\omega_0)\,\bm{\dip}\,\sigma_{ge}^{j}\,,
\end{equation}
where we assume that all emitter dipoles are equally oriented, with dipole moment $\bm{\dip}=|\bm{\dip}|\hat{\dip}$ and resonant frequency $\omega_0$. Furthermore, provided that the Green's tensor is soley given by the far-field contribution of the free-space one:
\begin{equation} \label{eq:G_0(R,rj,omega_0)}
    \dyad{G}_0(\RR,\rr_j,\omega_0)=\frac{e^{ik_0(|\RR|-\hat{R}\rr_j)}}{4\pi|\RR|} \left[\dyad{I}-\frac{\RR\otimes\RR}{|\RR|^2} \right],
\end{equation}
where $k_0=2\pi/\lambda_0$ with $\lambda_0$ the wavelength of the emitters' transition, it can be shown that:
\begin{equation}
    \bra{\Psi(t)}\mathbf{E}^{-}(\RR)\mathbf{E}^{+}(\RR)\ket{\Psi(t)} =  
    \left(\frac{\mu_0\omega_0^2|\bm{\dip}|}{4\pi|\RR|}\right)^2 
    f(\hat{R},\hat{\dip}).
\end{equation}
Here, we have used the overall-wavefunction ansatz given by Eq.~\eqref{eq:overall_wavefun}. The temporal dependence is hidden in the form factor $f(\hat{R},\hat{\dip})$ through the emitter populations $C_j(t)$:
\begin{equation}
    f(\hat{R},\hat{\dip}) = [(\hat{R}\times\hat{\dip})\times\hat{R}]^2
    \sum_{j,j^\prime=1}^{N}C_j^*(t)C_{j^\prime}(t)\,
    e^{ik_0\hat{R}\,\rr_{jj^\prime}},
\end{equation}
with $\hat{R}=\cos\theta\cos\phi\,\mathbf{\hat{e}}_\perp+\cos\theta\sin\phi\,\mathbf{\hat{e}}_\parallel+\sin\theta\,\mathbf{\hat{e}}_z$ and $\rr_{jj^\prime}=\rr_{j}-\rr_{j^\prime}$. In Fig.~\ref{fig:Fig4}(b), we plot precisely this form factor for different times provided that the photonic Weyl environment is prepared in the same phase than in Fig.~\ref{fig:Fig3} and Fig.~\ref{fig:Fig2}(c), showing again the emergence of the Fermi arc signatures in reciprocal space.

 \section{Photonic Fermi arc surface states as robust quantum links~\label{sec:harnessing}}

 \begin{figure*}[t]
   \centering
    \includegraphics[width=17.8cm]{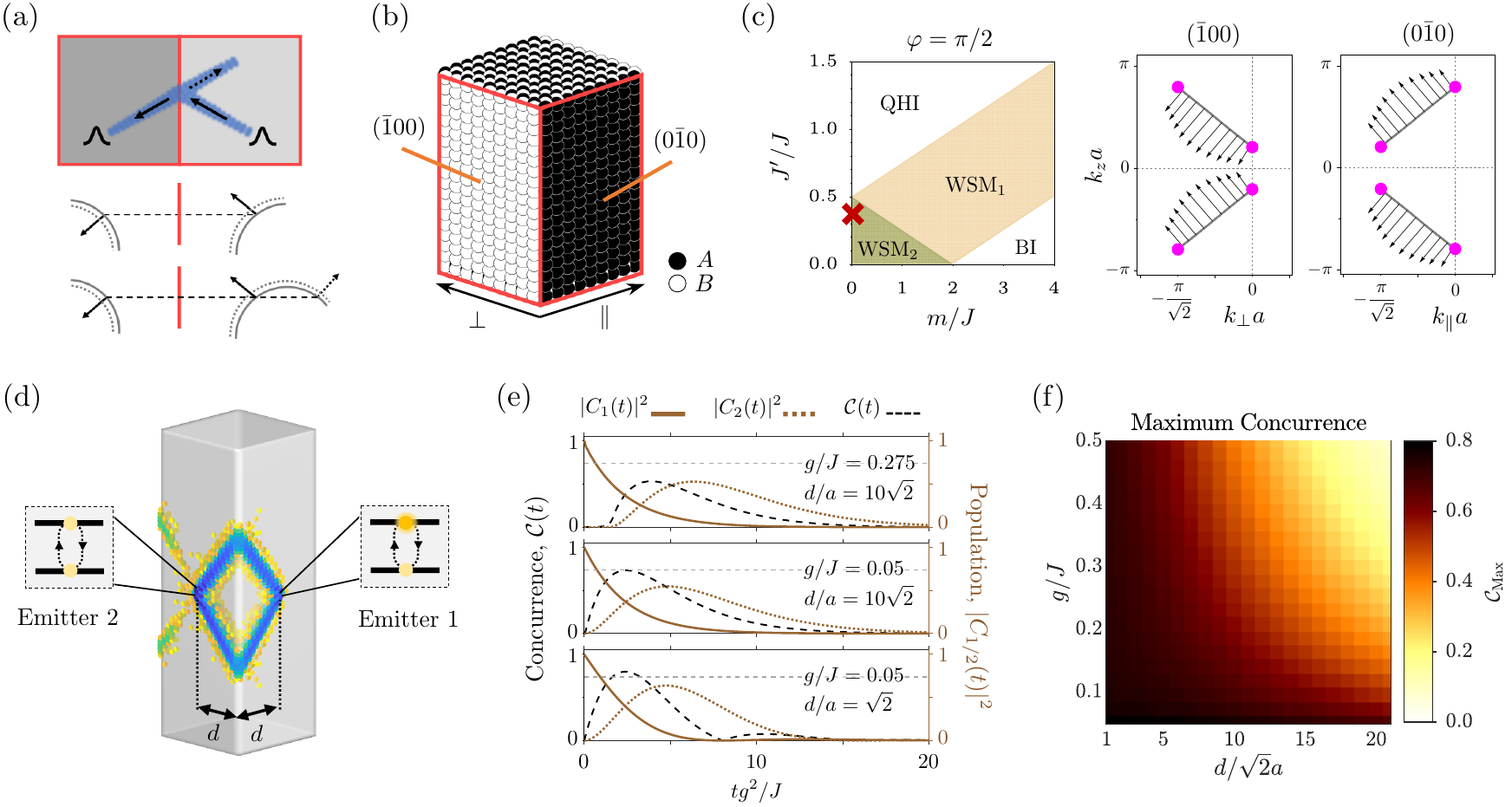}
       \caption{{\bf Utilization of the surface Weyl states as an ideal quantum link by exploiting the negative refraction of the photonic excitation at the system's hinges.} 
       (a)~Single mode picture of the negative refraction process occurring in the surface of a Weyl system. Middle and bottom panels depict the equifrequency contours associated to a system without and with back-propagation, respectively. 
       (b)~Boundary conditions required to produce negative refraction in the hinge formed between the $(0\bar{1}0)$ and the $(\bar{1}00)$ facets. 
       (c)~Phase diagram of the lattice model for $\varphi=\pi/2$ [see definition in Eq.~\eqref{eq:d(k)}]. Inset shows the Fermi arcs corresponding to the $(\bar{1}00)$ and $(0\bar{1}0)$ faces provided that the bath is prepared in configuration marked in the phase diagram with a red cross. Black arrows represent the group velocities associated to a set of selected $\kk$-points belonging to the Fermi arcs in arbitrary units.
       (d)~Investigated scenario in which two quantum emitters coupled to adjacent facets of the Weyl bath and separated some distance $d$ from the corner formed between the two considered facets. 
       (e)~Population dynamics and concurrence for different values of the light-matter coupling strength and distance between the emitters. 
       (f)~Maximum concurrence achieved as a function of the distance between emitters and the light-matter coupling strength. 
       }
 \label{fig:Fig5}
 \end{figure*}

After having shown in the previous section that emitters at the edges couple efficiently to the topological surface modes associated to the Fermi arcs, here we illustrate how to harness them to induce robust quantum links between the emitters. For that, we exploit one of the most striking features of such surface modes, that is, that they can lead to \textit{negative refraction} (NR) at the hinge that separates two different facets of the Weyl system \cite{He2018a}. Such effect, predicted originally by Veselago~\cite{Veselago1968} between materials with different ``rightnesses'', is triggering a lot of theoretical and experimental activity~\cite{He2018a, Yang2019, Deng2020, Chen2020b, Chen2020c, Zheng2021, Ukhtary2017, Yang2021, Yang2022, Tchoumakov2022} because of its potential uses, for example, to obtain perfect lensing~\cite{Pendry2000}. However, most of the applications so far have focused on the (semi)-classical regime. Here, we show how to exploit such phenomena to obtain a robust quantum link between emitters that can be harnessed for both perfect quantum state transfer~\cite{Cirac1997} and to induce maximal entanglement between the emitters in several ways. To show that, we divide this section in three parts:
\begin{itemize}
    \item In section~\ref{subsec:optimizeNR}, we first determine the system's parameters to enable NR in the Weyl system. Then, we optimize the bath configuration to maximize the coupling between emitters coupled to its edges. 
 
    \item Then, in section~\ref{subsec:chiralchannel} we consider the collective dynamics of two emitters placed at consecutive facets of a large enough system so that revival effects in the initially excited emitter do not occur. In that case, we demonstrate that, under certain conditions, the chiral propagation of the surface modes together with its NR at the hinge make that the photonic excitations behave as a perfect 1D, chiral channel~\cite{Lodahl2017}.
    
    \item Finally, in section~\ref{subsec:cavity} we consider the opposite limit where multiple refractions occur at the system's hinges. There, we show how, engineering the system properly, the light emitted from an emitter can arrive to the same point, forming effective 1D cavity modes. We demonstrate that such effective cavity modes induce perfect coherent exchanges between the emitters that can maximally entangle them.
\end{itemize}

\subsection{Optimizing Weyl system for negative refraction~\label{subsec:optimizeNR}}

In the discussion accompanying Fig.~\ref{fig:Fig2}, we already report on how an emitter tuned to the Weyl frequency excite preferentially the surface modes associated to the Fermi arc dispersions. Within a single facet, due to the mirror symmetry of the bath, an emitter always launches its excitations on two channels with opposite vertical component of the group velocity [see Fig.~\ref{fig:Fig2}(a)]. Such chiral, multi-channel emission could be used, e.g., for multiplexing quantum information in different directions~\cite{Guo2021,Li2022}. In this section, however, we are interested in the possibility of refocusing these channels onto a second emitter at the contiguous edge [see Fig.~\ref{fig:Fig5}(b)]. For that, we exploit the NR occurring at the system's hinges~\cite{He2018a, Yang2019, Deng2020, Chen2020b, Chen2020c, Zheng2021}.

A schematic view of the refraction process in the interface formed between two adjacent facets of the Weyl bath is depicted in Fig.~\ref{fig:Fig5}(a), where we represent an incoming Bloch state approaching from the right to one of the lattice's hinges. The group velocity of the targeted mode is given by a black arrow. At the interface, frequency and momentum parallel to the intersection are conserved~\cite{He2018a}, which implies that NR occurs if the tangential component of the group velocity changes its sign. Within this framework we can also identify the conditions to inhibit reflection, which entail the absence of resonant back-propagating modes in the first surface [compare the middle and bottom panels in Fig.~\ref{fig:Fig5}(a)]. A preliminary examination of the direction of growth of equifrequency contours associated to the $(0\bar{1}0)$ and the $(\bar{1}00)$ surfaces unveils that NR can be achieved between these two facets provided that the first one is composed by sites belonging to the $A$ sublattice whereas the second one is composed by sites belonging to the $B$ sublattice, as shown in Fig.~\ref{fig:Fig5}(b).

On top of that, to maximize the focusing onto a single spot, one must account for the fact that the group velocities characterizing the Bloch states of a given Fermi arc do generally feature slightly different propagation's directions. This causes each mode in the Fermi arc to undergo a distinct refraction angle, hindering the focalization of the photonic component in the second surface. To circumvent this problem, we explore the configurations' space of $H_\mathrm{B}$, examining different values of the parameter $\varphi$ defined in Eq.~\eqref{eq:d(k)}. We find that the optimal situation is obtained by choosing $\varphi=\pi/2$, instead of the $\varphi=0$ value used for Figs.~\ref{fig:Fig1}-\ref{fig:Fig3}. The phase diagram as a function of $m$ and $J^\prime$ for this value of $\varphi=\pi/2$ is displayed in Fig.~\ref{fig:Fig5}(c). The Fermi arcs of the $(0\bar{1}0)$ and the $(\bar{1}00)$ facets for the case $m/J=0$ and $J^\prime/J=0.4$ are plotted in the inset. The calculation of the group velocities associated to the $\kk$-points that conform these curves evidences an homogeneous distribution of the direction of propagation of the corresponding Bloch modes. To illustrate that, we plot the group velocities of some selected points in the Fermi arcs using black arrows. In what follows, we fixed this parameters' configuration and describe the emergence of the two working regimes explained in the introduction of this section.

To investigate both cases, we will employ the setup depicted in Fig.~\ref{fig:Fig5}(d), that is, considering two emitters separated at a distance $d$ from the intersection formed between the two studied facets. Besides, we will assume that the emitter at the $(0\bar{1}0)$ surface is excited, while the other one, in the $(\bar{1}00)$ surface, is initially in its ground state. After that, we let the system evolve freely and we track the populations of both the initially excited and the initially de-excited emitters which are given by $|C_1(t)|^2$ and $|C_2(t)|^2$, respectively. We also study the concurrence $\mathcal{C}(t)$ as a measure of the two-qubit entanglement~\cite{Wootters1998a}, which for these initial conditions can be shown to be given by~\cite{Maniscalco2008,Franco2013,Gonzalez-Ballestero2015}:
\begin{equation}
    \mathcal{C}(t)=2|C_1(t)C_2^*(t)|.
\end{equation}

There will be two relevant magnitudes that will distinguish the dynamical regimes that we will discuss in the next two sections. One is the expected decay time of the emitters that, within a Markovian regime, will be of the order $\tau\sim O(J/g^2)$. The other one is the time that an excitation will take to make a round-trip within the system, which will be of the order $T_\mathrm{R}\sim \frac{\ell}{Ja}$, with $\ell$ being the path length followed by the excitations, that will increase linearly with system size, and $Ja$ the typical group velocity that can be obtained in this system. As we will show below, the behaviour will be very different when $\tau\gg (\ll) T_\mathrm{R}$.

\subsection{Fermi arc surface states as perfect 1D, chiral channels~\label{subsec:chiralchannel}}

Let us first consider the limit of very large systems, that is, $\tau\ll T_R$. This means that the photons will decay from the emitters completely before any re-excitation can occur. This leads to an effective non-unitary dynamics in the emitters, since all the population is eventually lost into the bath. We simulate that regime numerically by including local losses as imaginary energies in the $(010)$ and $(100)$ facets, which attenuates any photonic excitation that arrives to them.

In spite of the non-unitary nature of the dynamics, the refocusing of the two channels due to the NR at the hinge [see Fig.~\ref{fig:Fig5}(d)] makes that the excitations leaked by the first emitter can be absorbed by the second. This is more clear in Fig.~\ref{fig:Fig5}(e) where we plot the population dynamics of the first (second) emitter in solid (dotted) brown lines, together with the associated concurrence $\mathcal{C}(t)$ in dashed black, for several parameters. There, we observe how the refocusing of the surface modes can induce a transient entangled state between the emitters.
This is the effect known as spontaneous generation of entanglement, which has been recently studied in different contexts~\cite{Dzsotjan2010,Gonzalez-Tudela2011a,Gangaraj2015,Gangaraj2017,Li2019,Ukhtary2022}, including one-dimensional chiral waveguides~\cite{Gonzalez-Ballestero2015}.

In the latter work, it was shown that, for a perfect Markovian chiral quantum optical channel, the maximum transient entanglement that can be achieved is $2/e\approx 0.73$. This is the value that we mark with the horizontal, dotted, gray line in Fig.~\ref{fig:Fig5}(e) showing how indeed we can reach the maximum in the situations where the emission is chiral and quasi-1D. Interestingly, for the smallest distance shown (see lower panel), the concurrence can go even beyond the ideal limit. This is attributed to the fact that, in that case, the emission does not have space to feature a quasi-1D nature, yielding values slightly above it. Finally, to complete the characterization, in Fig.~\ref{fig:Fig5}(f) we make a contour plot of the maximum transient value of concurrence for different corner-to-emitter distances $d$ and $g/J$ ratios observing how one can approach the ideal limit for a wide range of configurations.

Beyond the intrinsic interest of such spontaneous generation of entanglement, the most important value of the results illustrated by Figs.~\ref{fig:Fig5}(d-f) is that they prove that Fermi arc surface modes behave as a perfect 1D channel. 
This opens up the use of all the machinery developed for such systems~\cite{Lodahl2017}. For example, one can increase the value of entanglement by coherently driving the emitters with staggered frequencies~\cite{Ramos2014a,Pichler2015a,Ramos2016a}. In those works, it was shown how the combination of staggered drivings plus chiral quantum optical channels can lead to the formation of many-emitters entangled steady-states, where the concurrence can approach its maximum value, i.e., $\mathcal{C}(t\rightarrow \infty)=1$. Beyond two-level emitters, it is also well-known that, by using $\Lambda$-transitions controlled by Raman lasers, chiral quantum channels can be used to obtain deterministic quantum state transfer~\cite{Cirac1997}.

 \begin{figure}[t]
   \centering
    \includegraphics[width=\columnwidth]{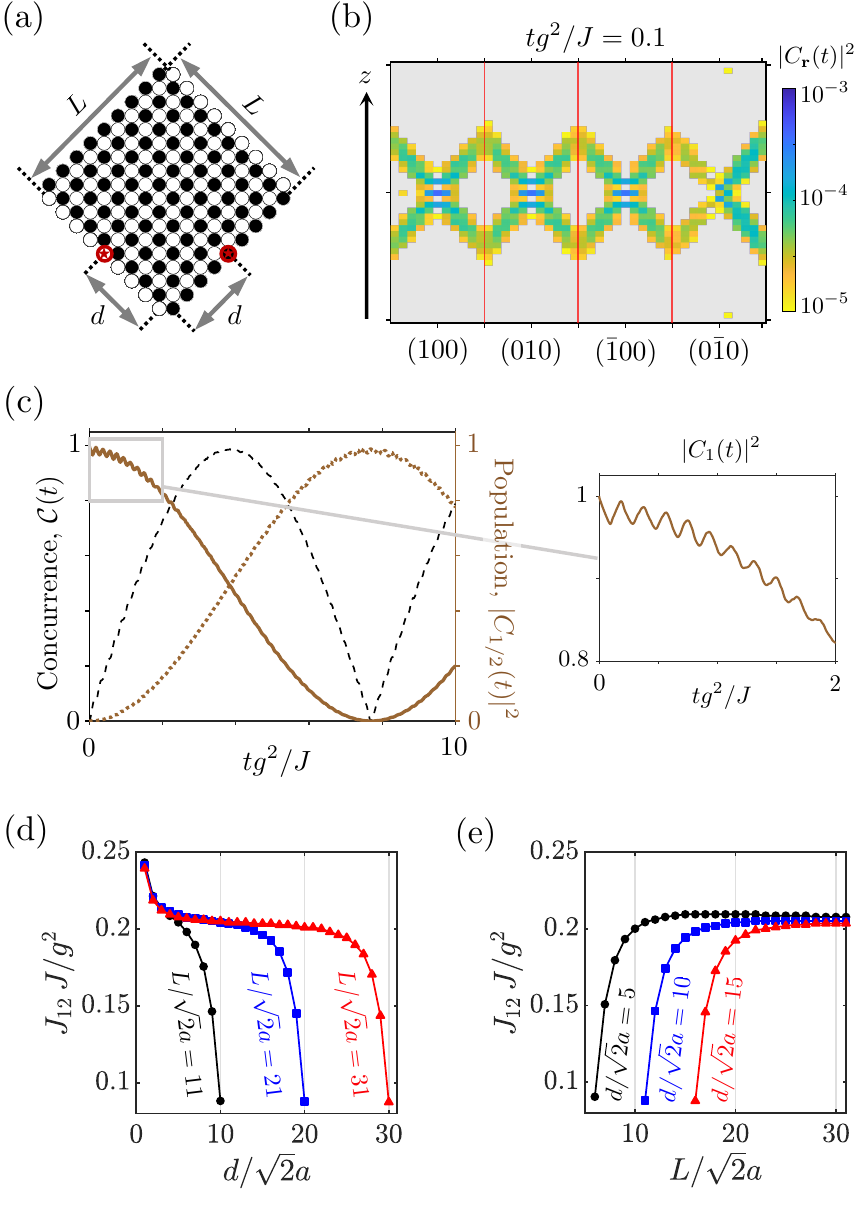}
       \caption{{\bf Emitters' dynamics and concurrence in the small system's size limit. } 
        (a) Top view of the finite lattice model employed to generate the effective cavity. Adjacent facets are composed by sites belonging to different sublattices ($A$ and $B$). 
        By imposing equally sized faces, we ensure the efficient revival of the considered emitters.  
        (b) Photonic population of the bosonic modes belonging to the four different facets in the lattice model for some selected de-excitation time.
        (c) Temporal evolution of the initially excited (solid brown) and initially de-excited (dashed brown) emitters. Dashed black line denotes the calculated concurrence. Inset shows a detailed view of the small oscillations observed at short times which are related to the time that the excitation spends to perform a complete round trip.
        (d) Oscillation frequency $J_{12}$ as a function of the corner-to-emitter distance $d$, for three different system's sizes $L/\sqrt{2}a=11,21$ and $31$ (black dots, blue squares, red triangles).
        (e) Oscillation frequency $J_{12}$ as a function of the system's size $L$, for three different corner-to-emitter distances $d/\sqrt{2}a=5,10$ and $15$ (black dots, blue squares, red triangles).
        }
 \label{fig:Fig6}
 \end{figure}

\subsection{Coherent exchanges induced by effective cavity modes. ~\label{subsec:cavity}}

In the small system's size limit, that is, when $T_R\ll \tau$, the occurrence of revivals due to the re-excitation of the emitters can not be avoided.
Rather than being a hindrance, we will show now how NR can turn these revivals into a resource. %
The key point is that, if one designs the system appropriately, as depicted in Fig.~\ref{fig:Fig6}(a), the photonic excitations circulate around the system arriving, eventually, to the position of the emitter from where they were launched. Again, this is possible due to the NR taking place at the system's hinges which guides the propagating ray through a confined, braid-shaped path. An example of that emission pattern is shown in Fig.~\ref{fig:Fig6}(b) where we plot a snapshot of the bath population in the facets for a particular time, showing the traces of the focusing and refocusing of light.

Remarkably, this light behaviour leads to a radically different dynamics of the emitters. An example of that is shown in Fig.~\ref{fig:Fig6}(c), where we plot the excited state population of the initially excited (de-excited) emitter in solid (dotted) brown. We observe that, differently from the previous situation, the emitters feature perfect coherent exchanges at a new frequency that we denote as $J_{12}$. We also observe small amplitude oscillations, which are zoomed in the right inset panel, whose frequency can be directly linked to the time it takes to the photonic excitations to perform a complete round-trip. Interestingly, the perfect coherent oscillations at frequency $J_{12}$ allows the emitter to reach the maximal entanglement of $\mathcal{C}(t)\approx 1$ in the transient regime (see dashed black line), overcoming the limitations of the chiral dissipative channels~\cite{Gonzalez-Ballestero2015}. 

The intuition on why these coherent oscillations appear is that, thanks to the NR, the photonic excitation undergo a closed loop, creating an effective one-dimensional cavity which is able to transfer excitations off-resonantly between the emitters~\cite{Ritsch2013a}.
To confirm this intuition and obtain further insight, we plot, in Fig.~\ref{fig:Fig6}(d), the frequency of the oscillations $J_{12}$ as a function of the corner-to-emitter distance for several system sizes. There, we observe that, disregarding the finite size effects that appear when the emitters are located close to the system's hinges, $J_{12}$ tends to be constant with the distance. This is more clear in the plateau obtained for $J_{12}$ for the larger system size (in red triangles). Note that this is what is expected for off-resonant cavity couplings since they typically mediate infinite-range interactions~\cite{Ritsch2013a}. Apart from that, in Fig.~\ref{fig:Fig6}(e) we plot the opposite situation, that is, we fix several distances between emitters, and study the dependence with system size. There, we observe how $J_{12}$ also tends to the same constant value for large system sizes. This can also be explained in terms of this effective cavity picture. Typically, off-resonant cavity couplings scale as $J_{12}\sim g_e^2/\Delta_e$, with $g_e$ being the coupling of the emitter to the effective cavity mode and $\Delta_e$ being its detuning. In these setups, the coupling strength always scales with the size of the cavity as $g_e\propto 1/\sqrt{\ell_\mathrm{c}}$. Besides, since the emitter's energy is fixed, the only dependence with system size in the detuning is that of the energy of the cavity modes, which are spaced by $\omega_n =n v_g/\ell_\mathrm{c}$~\cite{Ritsch2013a,Haroche2013a}. Thus, the dependence with system size in $J_{12}$ vanishes, which is why all lines of $J_{12}$ tend to a constant value in Fig.~\ref{fig:Fig6}(e).

\section{Conclusion~\label{sec:conclu}}

Summing up, we have characterized, for the first time, the behaviour of Fermi arc light-matter interfaces, discovering several remarkable phenomena. 
First, we have demonstrated that the studied platforms can be used to image the Fermi arcs in unconventional ways, by monitoring the free-space spontaneous emission of the considered emitters. More importantly, we have shown how to engineer the coupling to the surface modes so that they behave as a robust quantum channel in both a dissipative and coherent regime. Although, we illustrate this behaviour by studying the spontaneous formation of two-qubit entanglement between a pair of emitters, our results immediately open up their use for designing quantum state transfer protocols~\cite{Cirac1997} as well as for obtaining non-trivial entangled steady-states~\cite{Ramos2014a,Pichler2015a,Ramos2016a}, among other applications. The latter evidence the great potential of the Fermi arc surface states to be harnessed for quantum technological applications. Furthermore, we provide an alternative framework for the exploration of these topological states through their interaction with optically active emitters, uncovering phenomena with no analogue in fermionic systems, and which can be extended to study other topological surface states~\cite{Yang2019a,Devescovi2021,Devescovi2022}.

\section*{Acknowledgments}

I.G.-E. acknowledges financial support from Spanish Ministry for Science, Innovation, and Universities through FPU grant AP-2018-02748. A.G.-T. acknowledges financial support from the Proyecto Sinérgico CAM 2020 Y2020/TCS-6545 (NanoQuCo-CM), the CSIC Research Platform on Quantum Technologies PTI-001 and from Spanish project PID2021-127968NB-I00,
(MCIU/AEI/FEDER, EU). J.B.-A. and J. M. acknowledge financial support from the Spanish Ministry for Science, Innovation, and Universities through grants RTI2018-098452-B-I00 (MCIU/AEI/FEDER,UE) and MDM-2014-0377 (“Mar\'ia de Maeztu” programme for Units of Excellence in R\&D).

\begin{appendices}

\appendix
\section{Lattice model}~\label{apx:lattice}

\begin{figure*}[t]
	\centering
	\includegraphics[width=17.8cm]{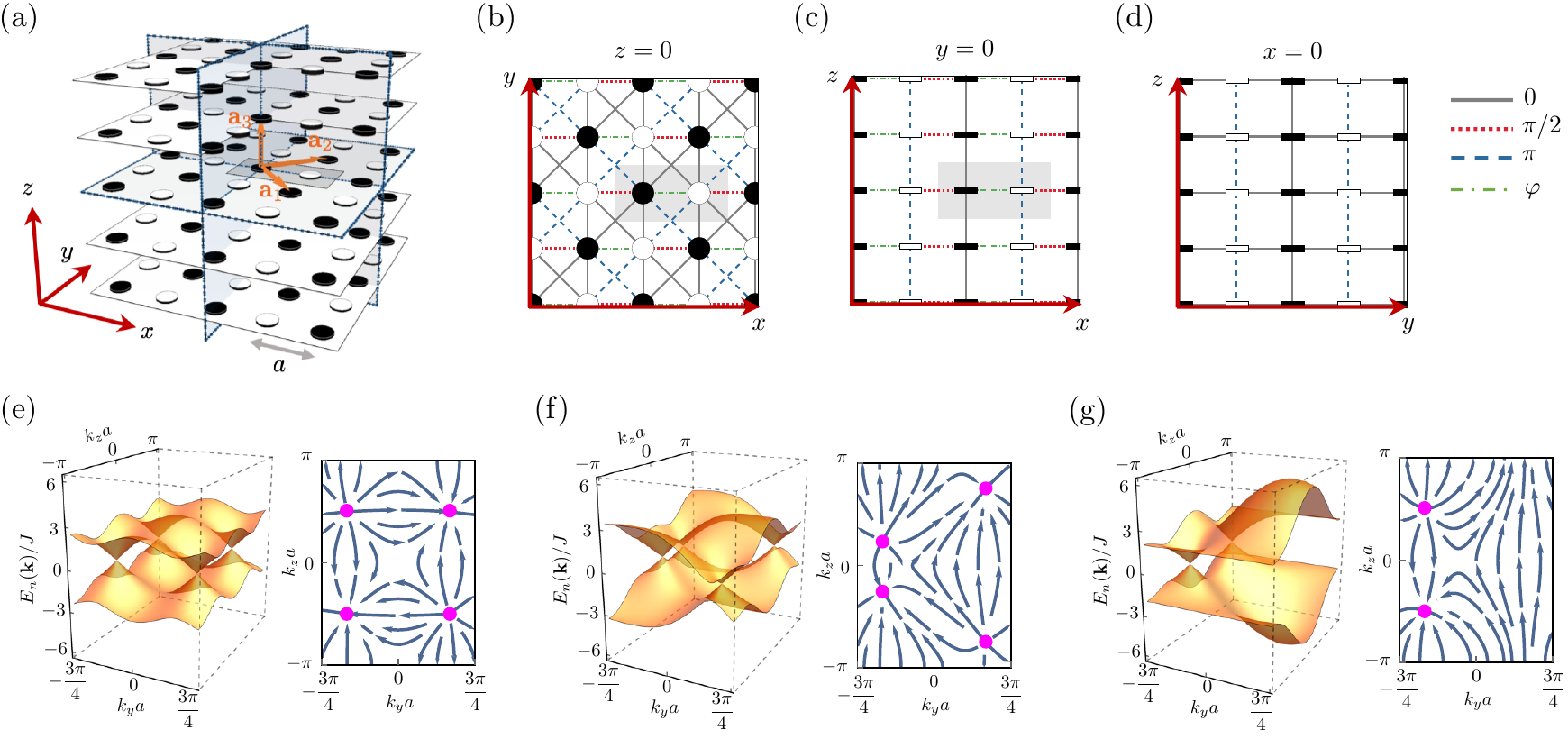}
	\caption{
	    {\bf Tight-binding model used to describe the photonic Weyl environment. } 
       (a) Three dimensional scheme of the employed lattice model. 
       (b-d) Top and side views of the lattice model and sketch of the employed hopping pattern.
       (e-g) Dispersion relation for fixed $k_xa=-\pi/4$ for the three configurations (I), (II) and (III) highlighted in the phase diagram of Fig.~\ref{fig:Fig1}(c). Insets show the Berry curvature associated to the corresponding cuts for the lower band $\bm{\Omega}^{-}(\kk)$.
	   }
	\label{fig:figS1}
\end{figure*}

In this appendix we provide further details on the lattice model that is used to mimic the photonic Weyl environment. 

The collection of bosonic modes that comprise the reservoir are positioned following a cubic geometry, where the corresponding lattice constant $a$ denotes the minimum separation between sites and is taken as the unit length throughout this section.
In order to keep the discussion as general as possible, arbitrary complex hoppings are assumed. This implies that, even in the simplest case in which only nearest-neighbors interactions are implemented, the discrete translation symmetry of the cubic arrangement is broken by the distribution of phases picked up by the excitation when it jumps among the different lattice's sites. Therefore, to take into consideration the magnetic symmetries of the system in all possible situations, a two-sites unit cell must be defined [without loss of generality we chose the unit cell as the gray shadowed area shown in Figs.~\ref{fig:figS1}(a-c)]. The lattice vectors are $\mathbf{a}_1=\ux+\uy$, $\mathbf{a}_2=\ux-\uy$ and $\mathbf{a}_3=\uz$, and they are depicted with orange arrows in Fig.~\ref{fig:figS1}(a). The position's ensemble spanned by each of the non-equivalent sites in the unit cell conform the $A$ and $B$ sublattices, which are represented in Figs.~\ref{fig:figS1}(a-d) by black and white cylinders, respectively. 
If periodic boundary conditions are imposed in all three spatial directions, the resulting lattice model captures well the dispersive properties of the modes belonging to the bulk of the system. Hereafter, we refer to this particular scheme as the bulk configuration. 

Alternatively, one can consider periodicity in only two spatial directions. This choice leads to the formalization of the slab configuration described in section~\ref{subsec:Weylbath} of the main text, which is used to investigate the edge states of the 3D bath. It is worth noting that, in this case, the direction in which the system is cut and the width of the slab are determined by the election of both, the surface lattice vectors $\bm{a}_1^s$ and $\bm{a}_2^s$, and the unit cell. The latter is, in general, different from the one defined in the  bulk configuration. More precisely, in the slab models, the number of non-equivalent sites belonging to the unit cell, i.e. the number of sublattices $N_s$, is usually larger than in the bulk configuration. 

Irrespective of the selected boundary conditions, the bath Hamiltonian $H_\mathrm{B}$ can be presented as the sum of four contributions:
\begin{equation}
    H_\mathrm{B}=H_{\mathrm{os}}+H_{xy}+H_{z}+H_{xy}^{\prime},
\end{equation}
which we express now using a basis of localized bosonic modes. The first term is given by: 
\begin{align}
H_{\mathrm{os}} =
& \sum_{\rr\in A}(\varepsilon-m)\bc{\rr}\bd{\rr}+
  \sum_{\rr\in B}(\varepsilon+m)\bc{\rr}\bd{\rr},
\end{align}
where $\varepsilon$ denotes the bare onsite energy of the bosonic modes and $m$ is a staggered mass term that produces an onsite energy off-set between modes belonging to different sublattices. The second and third terms represent the nearest-neighbors intra- and inter-layer interactions, respectively. They read as:
\begin{align}
\begin{split} 
H_{xy} = 
- & \sum_{\rr\in A} \J{}{1}\bc{\rr}\bd{\rr+\ux}+\J{}{2}\bc{\rr}\bd{\rr-\ux}+\cdots 
\\
\phantom{-} & \phantom{\sum_{\rr\in\mathrm{A}}} \J{}{3}\bc{\rr}\bd{\rr+\uy}+\J{}{4}\bc{\rr}\bd{\rr-\uy}+\mathrm{H.c.},
\end{split}
\\[1.0ex]
H_{z} = 
- & \sum_{\rr\in A}\left(\J{}{5}\bc{\rr}\bd{\rr+\uz}+\mathrm{H.c.}\right)
-   \sum_{\rr\in B}\left(\J{}{6}\bc{\rr}\bd{\rr+\uz}+\mathrm{H.c.}\right),
\end{align}
where $\J{}{j}=|\J{}{j}|e^{i\varphi_{j}}$ is a complex hopping matrix element. 
The last term accounts for some additional intra-layer next-nearest-neighbours interaction which is included to enrich the phase space of the model. It is described by the following tight-binding Hamiltonian:
\begin{equation}
    \begin{split} 
    H_{xy}^{\prime} = 
    - & \sum_{\rr\in A}\left(\J{}{7}\bc{\rr}\bd{\rr+(\ux+\uy)}+\J{}{8}\bc{\rr}\bd{\rr+(\ux-\uy)}+\mathrm{H.c.}\right)\cdots
    \\ 
    - & \sum_{\rr\in B}\left(\J{}{9}\bc{\rr}\bd{\rr+(\ux+\uy)}+\J{}{10}\bc{\rr}\bd{\rr+(\ux-\uy)}+\mathrm{H.c.}\right).
    \end{split}
\end{equation}

This real space representation of the bath Hamiltonian is specially useful to address finite sized lattices since it does not relay in any particular boundary condition. Conversely, if discrete translational invariance is imposed in at least one spatial direction, the problem can be simplified by introducing reciprocal space. 
To do that, we build a basis of Bloch-like orbitals, defined as follows:
\begin{equation}
    \ket{\phi_{\xi_n\kk}}=\bd{\xi_n\kk}^{\dagger}\ket{\mathrm{vac}},
\end{equation}
where $\ket{\mathrm{vac}}$ denotes the electromagnetic vacuum and $\bd{\xi_n\kk}^{\dagger}$ is the Fourier transformation of the bosonic operator $a_\rr^{\dagger}$, which is given by:
\begin{equation} \label{eq:FT_boson_operators}
    \bd{\xi_n\kk}^{\dagger}=\frac{1}{\sqrt{N_{\xi_n}}}\sum_{\rr\in\xi_n}e^{i\kk\rr}\bd{\rr}^{\dagger}.
\end{equation}
Here, $N_{\xi_n}$ is the total number of sites belonging to the $\xi_n$-th sublattice and $\kk$ is a vector in the first Brillouin zone. Using this relation, $H_\mathrm{B}$ can be rewritten as a quadratic form:
\begin{equation}\label{eq:H_B_cuadratic_form}
    H_\mathrm{B} = \sum_{\kk} A_{\kk}^\dagger \, \bar{H}_\mathrm{B}(\kk) \, A_{\kk},
\end{equation}
where we have defined the row operator $A_{\kk}^\dagger=(\bc{\xi_1\kk} \cdots \bc{\xi_{N_s}\kk})$ and the column operator $A_{\kk}=(A_{\kk}^\dagger)^\dagger$. 
Besides, we recognize $\bar{H}_\mathrm{B}(\kk)$ as the matrix of the bath Hamiltonian in the basis of Bloch-like orbitals, whose matrix elements read: $\left[\bar{H}_\mathrm{B}(\kk)\right]_{ij}=\bra{\phi_{\xi_i\kk}}H_\mathrm{B}(\kk)\ket{\phi_{\xi_j\kk}}$. This is just a $N_s\times N_s$ matrix that can be brought to its diagonal form by identifying the $\kk$-dependent unitary transformation $\bar{U}(\kk)$ that yields:
\begin{equation}
    \bar{D}_\mathrm{B}(\kk)=\bar{U}^\dagger(\kk)\bar{H}_\mathrm{B}(\kk)\bar{U}(\kk)
\end{equation}
where $\bar{D}_\mathrm{B}(\kk)$ represents a diagonal matrix. In particular we have that:
\begin{equation}\label{eq:H_B_cuadratic_form_diag}
    H_\mathrm{B} = \sum_{\kk} \tilde{A}_{\kk}^\dagger \, \bar{D}_\mathrm{B}(\kk) \, \tilde{A}_{\kk},
\end{equation}
where $\tilde{A}_{\kk}^\dagger=A_\kk^\dagger\bar{U}(\kk)=(\tilde{a}_{1\kk}^\dagger \cdots \tilde{a}_{N_s\kk}^\dagger)$ and $\tilde{A}_{\kk}=\bar{U}^\dagger(\kk) A_\kk$. It must be noticed that the non-zero elements of $\bar{D}_\mathrm{B}(\kk)$ account for the bands of the structured bath, i.e., $[\bar{D}_\mathrm{B}(\kk)]_{nn}\equiv E_{n}(\kk)$. This is, Eq.~\eqref{eq:H_B_cuadratic_form_diag} can be rewritten as:
\begin{equation}
    H_\mathrm{B}=\sum_{n\kk}E_{n}(\kk)\,\tilde{a}^\dagger_{n\kk}\tilde{a}_{n\kk}.
\end{equation}

Moreover, the Bloch state with associated band number $n$ and quasimomentum $\kk$ is given by:
\begin{equation} \label{eq:Bloch_state}
    \ket{\psi_{n\kk}}=\tilde{a}^\dagger_{n\kk}\ket{\mathrm{vac}},
\end{equation}
or, in the basis of Bloch-like orbitals:
\begin{equation}
    \ket{\psi_{n\kk}}=
    \sum_{i=1}^{N_s}u_{i n}\ket{\phi_{\xi_n\kk}},
\end{equation}
where $u_{ij}\equiv[\bar{U}(\kk)]_{ij}$ correspond to the matrix elements of the unitary transformation $\bar{U}(\kk)$.

Finally, we define the Berry curvature tensor of the $n$-th band $\Omega_{\mu\nu}^n(\kk)$, which can be computed by employing the Berry connection~\cite{Berry1984,Xiao2010}:
\begin{equation} \label{eq:Berry_curvature_tensor_Berry_connection}
    \Omega_{\mu\nu}^n(\kk) =
    \frac{\partial \mathcal{A}_\nu^n(\kk)}{\partial k_\mu}-
    \frac{\partial \mathcal{A}_\mu^n(\kk)}{\partial k_\nu},
\end{equation}
where $\bm{\mathcal{A}}^n(\kk)=i\bra{\psi_{n\kk}}\frac{\partial}{\partial\kk}\ket{\psi_{n\kk}}$.
Equivalently, using the Bloch-like orbitals basis, $\Omega_{\mu\nu}^n(\kk)$ can be calculated as follows (omitting the $\kk$ dependencies for clarity):
\begin{equation} \label{eq:Berry_curvature_tensor}
\begin{split} 
    \Omega_{\mu\nu}^n = i\sum_{n^{\prime}\neq n} 
    & \frac{
    (\bm{u}_n^{*}\,\frac{\partial \bar{H}_\mathrm{B}}{\partial k_\mu}\,\bm{u}_{n^\prime})\,
    (\bm{u}_{n^\prime}^{*}\,\frac{\partial \bar{H}_\mathrm{B}}{\partial k_\nu}\,\bm{u}_{n})}
    {(E_n-E_{n^\prime})^2} -(\nu\leftrightarrow\mu),
\end{split}
\end{equation}
where $\bm{u}_{n}\equiv(u_{1n},u_{2n},\cdots,\,u_{Nn})^T$ is a $\kk$-dependent vector representing the $n$-th column of the unitary transformation $\bar{U}(\kk)$ that diagonalizes $\bar{H}_\mathrm{B}(\kk)$.
We remark that the derivation of the band structure and Berry curvature tensor discussed above is valid for both the bulk and slab configurations. In the following we study each case separately.

\subsection{Bulk configuration}

When periodic boundary conditions are assumed in all three spatial directions,  $\bar{H}_\mathrm{B}(\kk)$ is given by a $\kk$-dependent $2\times2$ matrix since, in this case, the unit cell contains two non-equivalent sites, i.e., $N_s=2$. In particular, they correspond to the $A$ and $B$ sublattices. As in the real space representation, we separate $\bar{H}_\mathrm{B}(\kk)$ into four different contributions:
\begin{equation} \label{eq:Bloch_Hamiltonian_bulk}
    \bar{H}_\mathrm{B}(\kk)=\bar{H}_{\mathrm{os}}+\bar{H}_{xy}+\bar{H}_{xy}^{\prime}+\bar{H}_{z},
\end{equation}
where each contribution reads:
\begin{widetext}
\begin{align} 
\bar{H}_{\mathrm{os}} = & \;\varepsilon\mathds{1}-\stgmas\sigma_z, \label{eq:bar_H_os}
\\[1.0ex]
\begin{split} 
\bar{H}_{xy} = 
    & -\left[\right.|t_1|\cos(k_x+\varphi_{1})+|t_2|\cos(k_x-\varphi_{2})+|t_3|\cos(k_y+\varphi_{3})+|t_4|\cos(k_y-\varphi_{4})\left.\right]\sigma_x \cdots
    \\[0.5ex] 
    & +\left[\right.|t_1|\sin(k_x+\varphi_{1})-|t_2|\sin(k_x-\varphi_{2})+|t_3|\sin(k_y+\varphi_{3})-|t_4|\sin(k_y-\varphi_{4})\left.\right]\sigma_y,
\end{split} \label{eq:bar_H_xy}
\\[1.0ex] 
\bar{H}_{z} = 
    & -|t_5|\cos(k_z+\varphi_5)\,(\mathds{1}+\sigma_z)-|t_6|\cos(k_z+\varphi_6)\,(\mathds{1}-\sigma_z), \label{eq:bar_H_z}
\\[1.0ex]
\begin{split} 
\bar{H}_{xy}^{\prime} = 
    & -\left[\right.|t_7|\cos(k_{+}+\varphi_7)+|t_8|\cos(k_{-}+\varphi_8)\left.\right](\mathds{1}+\sigma_z)
      -\left[\right.|t_9|\cos(k_{+}+\varphi_9)+|t_{10}|\cos(k_{-}+\varphi_{10})\left.\right](\mathds{1}-\sigma_z).
\end{split} \label{eq:bar_Hprime_xy}
\end{align}
\end{widetext}
Here, we have defined $k_{\pm}=k_x \pm k_y$ and we have conveniently introduced the identity, $\mathds{1}$, and the Pauli matrices, $\bm{\sigma}=(\sigma_x,\sigma_y,\sigma_z)$. 
It is worth noting that, for this two-band model, $\bar{H}_\mathrm{B}(\kk)$ can be compactly express as follows:
\begin{equation} \label{eq:H_B_two-bands}
    \bar{H}_\mathrm{B}(\kk)=d_0(\kk)\mathds{1}+\mathbf{d(\kk)}\bm{\sigma},
\end{equation}
with $d_0(\kk)$ and the different components of $\mathbf{d(\kk)}$ being continuous functions of $\kk$. From Eqs.~\eqref{eq:bar_H_z} and \eqref{eq:bar_Hprime_xy}, is easy to see that, in order to get a type I semimetal, one must set $|\varphi_5-\varphi_6|=|\varphi_7-\varphi_9|=|\varphi_8-\varphi_{10}|=\pi$ and $|t_5|=|t_6|$, $|t_7|=|t_9|$ and $|t_8|=|t_{10}|$. If those conditions are imposed, one gets $d_0(\kk)=\varepsilon$, which we identify with the Weyl frequency, i.e., $\varepsilon\equiv\omega_W$. The band dispersion of the upper and lower bands are given by $E_{\pm}(\kk)=\omega_W\pm|\mathbf{d}(\kk)|$ which, provided that the system features a semimetallic phase, implies that the position of the Weyl nodes can be obtained by solving $|\mathbf{d}(\kk)|=0$. Besides, since the reciprocal space is three-dimensional, the topological properties of the bath are encoded in the Berry curvature vector $\bm{\Omega}^n(\kk)$, whose components are related to the Berry tensor by $\Omega_{\mu\nu}^n(\kk)=\bm{\epsilon}_{\mu\nu\xi}[\bm{\Omega}^n(\kk)]_\xi$~\cite{Xiao2010}. The latter can be explicitly work out yielding:
\begin{equation} \label{eq:BerryCurvature_2LS}
    \Omega_{\mu\nu}^{\pm}(\kk) =
    \mp\frac{\mathbf{d}(\kk)}{2|\mathbf{d}(\kk)|^3}\,
    \left[
    \frac{\mathbf{d}(\kk)}{\partial k_\mu}
    \times
    \frac{\mathbf{d}(\kk)}{\partial k_\nu}
    \right].
\end{equation}
Calculating the flux of $\bm{\Omega}^n(\kk)$ over a surface enclosing a Weyl point one can determine its corresponding Chern number.

The simplified model employed in the main text can be recovered by particularizing Eq.~\eqref{eq:Bloch_Hamiltonian_bulk} for the set of parameters collected in Table~\ref{tab:table1}. The dispersion relation along the $k_xa=-\pi/4$ plane for the configurations marked by (I), (II) and (III) in the phase diagram of Fig.~\ref{fig:Fig1}(d) in the main text are represented in the panels (e), (f) and (g) of Fig.~\ref{fig:figS1}, respectively. Insets show the Berry curvature (blue arrows) and the Weyl points (pink dots) for each of the studied configurations.

\subsection{Slab configuration}

If the discrete translation symmetry is broken along some specific direction while maintained along the remaining ones, the resulting lattice features the form of a slab. A particular example is introduced in section~\ref{subsec:Weylbath} of the main text. In this case, the associated Bloch Hamiltonian is given by a $N_s \times N_s$ matrix featuring the following form:
\begin{equation}
\bar{H}_\mathrm{B}(\kk) =
\begin{tikzpicture}[baseline=(current bounding box.center)]
\matrix (m)[
    matrix of math nodes,
    left delimiter={(},
    right delimiter={),},
    inner sep=1pt,
    nodes={draw,inner sep=2pt},
    column sep=0.5pt,row sep=0.5pt
    ] 
    {|[inner sep=2mm]|\bar{h}_{AA} & |[inner sep=2mm]|\bar{h}_{AB} \\
     |[inner sep=2mm]|\bar{h}_{BA} & |[inner sep=2mm]|\bar{h}_{BB} \\};
     
\end{tikzpicture}
\end{equation}
where the $\bar{h}_{AA(BB)}$ block accounts for the interaction among sites belonging to the $A(B)$ sublattice, whereas the $\bar{h}_{AB}$ and $\bar{h}_{BA}=\bar{h}_{AB}^\dagger$ blocks describe the interactions between sites belonging to different sublattices. 
Even though the definition of the $A$ and $B$ sublattices is strongly related to the bulk configuration, we maintain such distinction in the construction of the matrix Hamiltonian for the slab configuration. The latter requires consider the proper ordering of the Fourier-transformed bosonic operators that conform the row $A_\kk^\dagger$ and column $A_\kk$ operators referred to in Eq.~\eqref{eq:H_B_cuadratic_form}.

The number of non-equivalent sites conforming the unit cell $N_s$ determines, in this case, the width of the slab and, therefore, the dimension of $\bar{H}_\mathrm{B}(\kk)$. The matrix elements for each of the considered blocks depends on the direction of the cut. For the example discussed in section~\ref{subsec:Weylbath}, each blocks is given by a tridiagonal matrix, where the non-zero matrix elements read:
\begin{align} \label{eq:Hbloch_slab}
    & \left[\bar{h}_{AA(BB)}\right]_{i,i} = -(+)m-[t_{5(6)}\,e^{ik_z}+t_{7(9)}\,e^{i\sqrt{2}k_\parallel}+\mathrm{H.c.}], 
    \\
    & \left[\bar{h}_{AA(BB)}\right]_{i,i+1} = -t_{8(10)},
    \\
    & \left[\bar{h}_{AA(BB)}\right]_{i+1,i} = -t_{8(10)}^*,
\end{align}
and
\begin{align}
    & \left[\bar{h}_{AB}\right]_{i,i} = -t_1\,e^{ik_\parallel/\sqrt{2}}-t_4\,e^{-ik_\parallel/\sqrt{2}}, 
    \\
    & \left[\bar{h}_{AB}\right]_{i,i+1} = 0,
    \\
    & \left[\bar{h}_{AB}\right]_{i+1,i} = -t_2\,e^{ik_\parallel/\sqrt{2}}-t_3\,e^{-ik_\parallel/\sqrt{2}}, 
\end{align}

We note that, for the investigated example, an odd number of non-equivalent sites are considered. Also, we chose the $(0\bar{1}0)$ and $(\bar{1}00)$ faces to be composed by sites belonging to the $A$ sublattice. This implies that the dimensions of the $\bar{h}_{AA}$ and the $\bar{h}_{BB}$ blocks are different, namely, we have that the latter are square matrices with dimensions $(N_s+1)/2$ and $(N_s-1)/2$, respectively.

Finally, we point out that the surface Berry curvature associated the the slab configuration can be computed using the general expression given by Eq.~\eqref{eq:Berry_curvature_tensor}. This is what we represent in the color maps of Figs~\ref{fig:Fig1}(e-g). More precisely, we plot the Berry curvature associated to the $n=17$ band of the considered slab configuration which, provided that we select a unit cell formed by $N_s=33$ non-equivalent sites, is identified as the edge band, i.e., $\Omega_\mathrm{eb}(\kk)\equiv\Omega_{\parallel z}^{n=17}(\kk)$ (see details in Appendix~\ref{apx:slabBC}).

\begin{table}[h]
\centering
\begin{tabular}{c c | c c}
\toprule
\multicolumn{2}{c|}{Nearest-neighbours} & \multicolumn{2}{c}{Next-nearest-neighbours} \\ [0.8ex] 
\hline\hline
\midrule
Amplitude     & Phase                       & Amplitude             & Phase         \\ \hline
$|t_{1}|=J$ & $\varphi_{1}=\varphi$       & $|t_{7}|=J^\prime$  & $\varphi_{7}=0$   \\
$|t_{2}|=J$ & $\varphi_{2}=\frac{\pi}{2}$ & $|t_{8}|=J^\prime$  & $\varphi_{8}=\pi$ \\
$|t_{3}|=J$ & $\varphi_{3}=0$             & $|t_{9}|=J^\prime$  & $\varphi_{9}=\pi$ \\
$|t_{4}|=J$ & $\varphi_{4}=0$             & $|t_{10}|=J^\prime$ & $\varphi_{10}=0$  \\
$|t_{5}|=J$ & $\varphi_{5}=0$             &                     &                   \\
$|t_{6}|=J$ & $\varphi_{6}=\pi$           &                     &                   \\
\bottomrule
\end{tabular}%
\caption{Parameters employed to obtain the lattice model presented in the main text.}
\label{tab:table1}%
\end{table}%

\section{Topological characterization of the gapped phases.}~\label{apx:topo_phases}

The topological characterization of the gapped phases found in the phase diagram shown in Fig.\ref{fig:Fig1}(c) can be done using a dimensional reduction approach~\cite{Jiang2012,Delplace2012a}. The idea is to treat one of the quasi-momentum variables as a free parameter, such that the resulting matrix Hamiltonian corresponds to a 2D system. In that case, the Chern number associated to the upper and lower bands is given by:
\begin{equation}
    C_{\pm} = 
    \frac{1}{2\pi}\int_{\mathrm{BZ}}d^2\kk\,\Omega^{\pm}(\kk).
\end{equation}
where $\Omega^{\pm}(\kk)$ denotes the Berry curvature of the effective model. In particular, by taking $k_z$ as the free parameter in the matrix Hamiltonian defined in Eq.~\eqref{eq:Bloch_Hamiltonian}, we have that $\Omega^{\pm}(\kk)\equiv\Omega_{xy}^\pm(\kk)$, where the RHS is explicitly given by Eq.~\eqref{eq:BerryCurvature_2LS}. Importantly, since $k_z$ is treated as a parameter, we must view the Chern number as a $k_z$-dependent quantity.
An alternative approach to calculate the Chern number consists in using the Brouwer degree of the mapping $\kk\rightarrow\hat{\mathbf{d}}(\kk)=\mathbf{d}(\kk)/|\mathbf{d}(\kk)|$, which leads for the following general expression \cite{Cayssol2021}:
\begin{equation}\label{eq:Brouwer_degree}
    C_\pm=\mp\frac{1}{2}\sum_{\kk_D}\chi(\kk_D)\,\mathrm{sign[d_z(\kk)]}
\end{equation}
where $\chi(\kk_D)=\pm1$ is the winding number of the Dirac point located at $\kk_D$.

\begin{figure}[t]
	\centering
	\includegraphics[width=8.7cm]{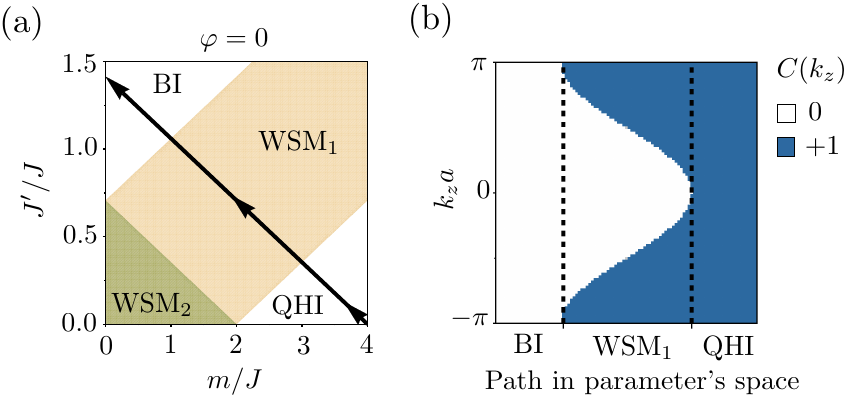}
	\caption{
       {\bf Topological characterization of the gapped phases in the phase diagram.}
       (a) Selected path in parameters' space to demonstrate the different topological nature of the band, and quantum Hall (anomalous) insulating phases (BI and QHI, respectively).
       (b) Chern number calculated for the set of parameters defined by the path highlighted in panel (a).
	   \vspace{-0.3cm}}
	\label{fig:FigS2}
\end{figure}

Next, to demonstrate the different topological nature of the phases in the phase diagram of Fig.\ref{fig:Fig1}(c), we calculate the $k_z$-dependent Chern number for the path in parameters' space marked with a black solid line in Fig.~\ref{fig:FigS2}(a). The components of the $\kk$-dependent vector $\mathbf{d}(\kk)$ for the studied case result from particularizing for $\varphi=0$ in Eq.~\eqref{eq:d(k)}, which yields:
\begin{equation}
    \begin{cases}
    d_x(\kk) = -\sqrt{2}J\cos(k_x-\frac{\pi}{4})-2J\cos(k_y),
    \\[0.2ex]
    d_y(\kk) = +\sqrt{2}J\cos(k_x-\frac{\pi}{4}),
    \\[0.2ex]
    d_z(\kk) = -m - 2J_z\cos(k_z) + 4J^{\prime}\sin(k_x)\sin(k_y),
\end{cases}
\end{equation}
where the corresponding Dirac points $\kk_D$ are obtained by solving $d_x(\kk_D)=d_y(\kk_D)=0$. By doing so, we found two Dirac points that we denote as:
\begin{equation}
    \mathbf{K}_\pm=(-\frac{\pi}{4},\pm\frac{\pi}{2}).
\end{equation}
Then, in order to calculate the winding number of the Dirac points, we expand the matrix Hamiltonian around $\mathbf{K}_\pm$:
\begin{align}
    & \bar{H}_\mathrm{B}(\kk\sim\mathbf{K}_+)\approx(-\sqrt{2}Jq_x+2Jq_y)\sigma_x+\sqrt{2}Jq_x\sigma_y
    \\[1.0ex]
    & \bar{H}_\mathrm{B}(\kk\sim\mathbf{K}_-)\approx(-\sqrt{2}Jq_x-2Jq_y)\sigma_x+\sqrt{2}Jq_x\sigma_y
\end{align}
where we have defined $q_x=k_x+\pi/4$ and $q_y\mp\pi/2$. Therefore, the velocity tensors of the Dirac cones are defined as follows:
\begin{equation}
    \bar{v}_\pm=
    \left(\begin{array}{cc}
        -\sqrt{2}J & +\sqrt{2}J\\
        \pm 2J     & 0
    \end{array}\right),
\end{equation}
and the winding numbers are given by $\chi(\mathbf{K}_\pm)=\mathrm{sign}[\mathrm{det}(\bar{v}_\pm)]$, which yields:
\begin{equation}
    \chi(\mathbf{K}_\pm)=\mp1
\end{equation}
Finally, using the formula given by Eq.~\eqref{eq:Brouwer_degree}, we obtain:
\begin{equation}
\begin{split}
    C_\pm(k_z) = \mp \frac{1}{2} \biggl(
    & \mathrm{sign}\left[-m - 2J_z\cos(k_z) + 2\sqrt{2}J^\prime\right] - 
    \\
    & \mathrm{sign}\left[-m - 2J_z\cos(k_z) - 2\sqrt{2}J^\prime\right]
    \biggr)
\end{split}
\end{equation}

The calculated Chern number corresponding to the lower band $C_{-}(k_z)$ is plotted in Fig.~\ref{fig:FigS2}(a) as a function of $k_z$ and the path followed in the parameter space. As seen, $C_{-}(k_z)=0$ for all $k_z$'s when the system is in the $\mathrm{BI}$ phase, and $C_{-}(k_z)=+1$ for all $k_z$'s when the system is in the $\mathrm{QHI}$ phase. When the system is in the $\mathrm{WSM_1}$ phase the calculated topological invariant is not homogeneous along the considered $k_z$ range.

\section{Berry curvature of the Weyl semimetal slab}~\label{apx:slabBC}

One of the fundamental aspects of Weyl semimetals is the fact that Weyl points behave as monopoles of the Berry curvature. 
A naturally arising question is how this unique condition translates into the geometrical properties of the Hilbert space describing the surface states of an open Weyl environment.
It turns out that the surface Berry curvature (i.e., the Berry curvature calculated over the surface Brillouin zone) of these class of topological reservoirs presents some interesting features. For example, the emergence of ``hot lines'' along which the Berry curvature exhibits a divergent behaviour has been demonstrated, very recently, using a continuous model~\cite{Wawrzik2021}. Interestingly, these ``hot lines'' are predicted to strongly affect the the nonlinear Hall effect in electronic systems.

For completeness, here, we describe how we computed the Berry curvature of the bands supporting the Fermi arcs in the studied lattice scheme in the last part of section~\ref{subsec:Weylbath}.
We first note that the number of Bloch bands resulting from the diagonalization of $H_\mathrm{B}$ when the slab geometry is implemented coincides with the number of non-equivalent sites conforming the extended unit cell. If the later comprises an odd number of sites, the dispersion relation of the slab model will consist of two separated sets of bulk bands connected by a single edge band [see red surface in the inset of Fig.~\ref{fig:Fig1}(d)].
For the considered slab, the $k_\parallel$ and $k_z$ quasi-momentum components are well defined. Therefore, the surface Berry curvature can be calculated as follows~\cite{Berry1984,Haldane2004,Xiao2010}:
\begin{equation}
    \Omega^n(\kk) = 
    \frac{\partial\mathcal{A}^n_z(\kk)}{\partial k_\parallel} -
    \frac{\partial\mathcal{A}^n_\parallel(\kk)}{\partial k_z}
\end{equation}
where $\bm{\mathcal{A}}^n(\kk)$ is the Berry connection of the $n$-th band [see Eq.~\eqref{eq:Berry_curvature_tensor_Berry_connection} in Appendix \ref{apx:lattice}].

The color maps in Fig.~\ref{fig:Fig1}(e-g) display the Berry curvature corresponding to the edge band $\Omega_{\mathrm{eb}}(\kk)$, for the three different configurations marked in the phase diagram of Fig.~\ref{fig:Fig1}(d). The Berry curvature in the first case is always zero due to the sublattice-dependent structure of the Bloch states associated to the edge band in the slab model, but the distribution observed for the (II) and (III) configurations shows some regions for which the Berry curvature present a non-trivial value. These regions in the surface Brillouin zone are associated to areas in which the localization of the wavefunction changes drastically~\cite{Wawrzik2021}. 
The obtained results represent a novel instance of the unique Berry-curvature effects that are predicted to appear in the surface Brillouin zone of Weyl semimetals. In particular, we show that the ``hot lines'' of divergent Berry curvature studied in Ref.~\cite{Wawrzik2021} emerge also naturally in a discrete lattice model.

\section{Mapping between real and reciprocal space}~\label{apx:mapping}

In this section we formalize the mapping established between the real space propagation of the photonic excitation through the lattice sites and the Fermi arc representations illustrated in Figs.~\ref{fig:Fig2}(b-d). To do that, we study the photonic component of the overall-wavefunction ansatz given by Eq.~\eqref{eq:overall_wavefun}, which we identify as:
\begin{equation} \label{eq:Photonic_component_r}
    \ket{\Psi_\mathrm{ph}(t)}=\sum_\rr C_{\rr}(t)\,\bc{\rr}\,\ket{\mathrm{vac}},
\end{equation}
where $\ket{\mathrm{vac}}$ denotes the electromagnetic vacuum. 
More precisely, we aim to rewrite $\ket{\Psi_\mathrm{ph}(t)}$ as follows:
\begin{equation} \label{eq:Photonic_component_k}
    \ket{\Psi_\mathrm{ph}(t)}=\sum_\kk C_{n\kk}(t)\,\tilde{a}_{n\kk}^\dagger\,\ket{\mathrm{vac}},
\end{equation}
where our main goal is to determine the form of the $C_{n\kk}(t)$ coefficients. These can be physically interpreted as the projections of the overall system state at instant $t$ over the Bloch modes that diagonalize $H_\mathrm{B}$, therefore, $|C_{n\kk}(t)|^2$ represents the population of Bloch modes associated to the band $n$ and quasi-momentum $\kk$.
We start by introducing an additional label to specifically account for the sublattice degree of freedom in Eq.~\eqref{eq:Photonic_component_r}:
\begin{equation}
    \ket{\Psi_\mathrm{ph}(t)}=\sum_{\xi_n}\sum_{\rr\in\xi_n} C_\rr(t)\,\bc{\rr}\,\ket{\mathrm{vac}},
\end{equation}
which, utilizing the Fourier transformation of the bosonic operators defined in Eq.~\eqref{eq:FT_boson_operators}, can be rewritten as:
\begin{equation}
    \ket{\Psi_\mathrm{ph}(t)}=\sum_{\kk}\,
    A^\dagger_\kk\,\mathbf{V}(\kk,t)
    \ket{\mathrm{vac}},
\end{equation}
where $A^\dagger_\kk$ stands for the row vector operator introduced in Eq.~\eqref{eq:H_B_cuadratic_form} and we have defined the column vector $\mathbf{V}(\kk,t)$, whose $n$-th component reads:
\begin{equation}
    \left[\mathbf{V}(\kk,t)\right]_n=\frac{1}{\sqrt{N_{\xi_n}}}\sum_{\rr\in\xi_n} C_\rr(t)e^{-i\kk\rr}
\end{equation}

Finally, we introduce the unitary transformation $\bar{U}(\kk)$ that diagonalizes the Bloch Hamiltonian $\bar{H}_\mathrm{B}(\kk)$ [see Eq.~\eqref{eq:H_B_cuadratic_form_diag}], to express $\ket{\Psi_\mathrm{ph}(t)}$ as:
\begin{equation}
    \ket{\Psi_\mathrm{ph}(t)}=\sum_{\kk}\,
    \underbrace{A^\dagger_\kk\bar{U}(\kk)}_{\equiv\tilde{A}_\kk^\dagger}\,
    \underbrace{\bar{U}^\dagger(\kk)\mathbf{V}(\kk,t)}_{\equiv\tilde{\mathbf{V}}(\kk,t)}
    \ket{\mathrm{vac}},
\end{equation}
which yields:
\begin{equation}
    \ket{\Psi_\mathrm{ph}(t)}=\sum_{\kk}\,
    [\tilde{\mathbf{V}}(\kk,t)]_n\,\tilde{a}_{n\kk}^\dagger
     \ket{\mathrm{vac}},
\end{equation}
where we identify $C_{n\kk}(t)=[\tilde{\mathbf{V}}(\kk,t)]_n$.

We would like to remark that the presented mapping procedure explicitly accounts for the sublattice structure of the bath. Furthermore, we claim that similar strategy can be employed even in the case in which more involved structure of the photonic environment is assumed, e.g., if additional ``internal'' degrees of freedom, such as orbital or spin degrees of freedom, are included.

\section{Time of flight}~\label{apx:time_of_flight}

In standard time of flight experiments a Bose-Einstein condensate trapped into an optical potential is released resulting into free expansion of the atomic cloud. The spatial density distribution of the ensemble is obtained via absorption imaging. In particular, after certain expansion time, $t_{ToF}$, the atoms are illuminated by a resonant laser beam. Part of the injected light is absorbed by the atomic cloud casting a shadow that is recorded in a CCD camera aligned with the laser direction. The ratio between the recorded intensity pattern, $I(x,y)$, and the profile obtained when the atomic cloud is absent, $I_0(x,y)$, is related to the column density, $\tilde{n}(x,y)$, which corresponds to the total density distribution, $\tilde{n}(\rr)$, integrated along the imaging direction. The total density distribution can be calculated as follows~\cite{Gerbier2005}:
\begin{equation} \label{eq:n_ToF(r)}
    \tilde{n}(\rr) = \left(\frac{m}{\hbar t_{ToF}}\right)^3 
    \left|w\left(\kk=\frac{m\rr}{\hbar t_{ToF}}\right)\right|^2 
    n\left(\kk=\frac{m\rr}{\hbar t_{ToF}}\right),
\end{equation}
where $m$ is the mass of the atoms in the ensemble and $w(\kk)$ is the Fourier transform of the on-site Warnnier function. Following~\cite{Kashurnikov2002}, we will ignore the latter function by formally setting it to unity. Then, we find that, when $\kk$ is restricted to the first Brillouin zone, the total density distribution is essentially proportional to the quasimomentum distribution, $n(\kk)$, which reads as:
\begin{equation} 
    n(\kk)=\sum_{j,j^\prime}e^{i\kk(\rr_j-\rr_{j^\prime})}\bra{\Psi(t)}a^\dagger_{\rr_j} a_{\rr_{j^\prime}}\ket{\Psi(t)},
\end{equation}
where $\ket{\Psi(t)}$ corresponds to the overall-system wavefunction at the released time.

\end{appendices}



%


\end{document}